\documentclass[journal]{IEEEtran}
\ifCLASSINFOpdf
\else
\fi

\usepackage{graphicx}
\usepackage{cite}

\usepackage{color,soul}

\usepackage{float}
\usepackage{color}
\usepackage{graphicx}
\usepackage{graphicx}
\usepackage{epstopdf}
\usepackage{amssymb,amsmath}
\usepackage{booktabs}
\usepackage{array}
\usepackage{multirow}
\usepackage{multicol}
\usepackage{hyperref}
\usepackage{comment}
\usepackage{newtxtext,newtxmath}
\usepackage[table,xcdraw]{xcolor}

\newcommand{\hermtranspose}{^\dagger}

\newcommand{\intensity}{{h}}
\newcommand{\reflectivity}{{x}}
\newcommand{\speckle}{{m}}
\newcommand{\logintensity}{\tilde{\intensity}}
\newcommand{\logreflectivity}{\tilde{\reflectivity}}
\newcommand{\logspeckle}{\tilde{\speckle}}
\newcommand{\covmat}{{\bf{\Sigma}}}

\newcommand{\diffvector}{{\boldsymbol k}}
\newcommand{\empcov}{{\bf C}}

\begin{document}
\markboth{\centering IEEE GEOSCIENCE AND REMOTE SENSING MAGAZINE, PREPRINT. PUBLISHED VERSION (2021): 10.1109/MGRS.2021.3121761}{}

%
\title{Image Restoration for Remote Sensing: Overview and Toolbox}

%
%
%

\author{Behnood~Rasti*,~\IEEEmembership{Senior~Member,~IEEE,}
        Yi~Chang*,~\IEEEmembership{Member,~IEEE,} Emanuele~Dalsasso*, Loic~Denis*, 
        and Pedram~Ghamisi*,~\IEEEmembership{Senior~Member,~IEEE}
\thanks{*All the authors have similar contributions in writing the manuscript.}
 \thanks{Behnood Rasti is with Helmholtz-Zentrum Dresden-Rossendorf, Helmholtz Institute Freiberg for Resource Technology, Machine Learning Group, Chemnitzer Straße 40, 09599 Freiberg, Germany; b.rasti@hzdr.de}
 \thanks{Yi Chang is with Science and Technology on Multispectral Information Processing Laboratory, School of Artificial Intelligence and Automation, Huazhong University of Science and Technology, Wuhan, China; owuchangyuo@gmail.com}
\thanks{Emanuele Dalsasso is with LTCI, Télécom Paris, Institut Polytechnique de Paris, Palaiseau; France, emanuele.dalsasso@telecom-paris.fr}%
\thanks{Loic Denis is with Univ Lyon, UJM-Saint-Etienne, CNRS, Institut d Optique Graduate School, Laboratoire Hubert Curien UMR 5516, F-42023, SAINT-ETIENNE, France; loic.denis@univ-st-etienne.fr}
\thanks{Pedram Ghamisi is with (1) Helmholtz-Zentrum Dresden-Rossendorf, Helmholtz Institute Freiberg for Resource Technology, Machine Learning Group, Chemnitzer Straße 40, 09599 Freiberg, Germany; (2) Institute of Advanced Research in Artificial Intelligence (IARAI), Landstraßer Hauptstraße 5, 1030 Vienna, Austria; p.ghamisi@gmail.com.}
}

\IEEEaftertitletext{\centering This is the pre-acceptance version, to read the final version published in 2021 in the IEEE Geoscience and Remote Sensing Magazine (IEEE GRSM), please go to: \textcolor{blue}{10.1109/MGRS.2021.3121761}\vspace{1\baselineskip} }

\maketitle

\begin{abstract}
Remote sensing provides valuable information about objects or areas from a distance in either active (e.g., RADAR and LiDAR) or passive (e.g., multispectral and hyperspectral) modes. The quality of data acquired by remotely sensed imaging sensors (both active and passive) is often degraded by a variety of noise types and artifacts. Image restoration, which is a vibrant field of research in the remote sensing community, is the task of recovering the true unknown image from the degraded observed image. Each imaging sensor induces unique noise types and artifacts into the observed image. This fact has led to the expansion of restoration techniques in  different paths according to each sensor type. This review paper brings
together the advances of image restoration techniques with particular focuses on synthetic aperture radar and hyperspectral images as the most active \textcolor{black}{sub-fields} of image restoration in the remote sensing community. We, therefore, provide a comprehensive, discipline-specific starting point for researchers at different levels (i.e., students, researchers, and senior researchers) willing to investigate the vibrant topic of data restoration by supplying sufficient detail and references. Additionally, this review paper accompanies a toolbox to provide a platform to encourage interested students and researchers in the field to further explore the restoration techniques and fast-forward the community. \textcolor{black}{The toolboxes are provided in https://github.com/ImageRestorationToolbox}.
\end{abstract}


%
\IEEEpeerreviewmaketitle

\section{Introduction}
\subsection{General Introduction}
Remote sensing (i.e., analyzing the immediate surface of the Earth using airborne or space-borne data) provides non-invasive techniques for target detection, analysis, and the observation of the Earth. Recent advances in remote sensing technologies and their variety open a broad range of applications related to Earth observation and monitoring. On the other hand, the complexity of remote sensing imaging technologies, together with the variety of noise sources and other nuisances (induced either by the environment or those technologies), make the interpretation of different data sources very challenging. 

Remote sensing data restoration attempts to recover an image from its corrupted version. The recovered image improves further analysis of the remote sensing images. Remote sensing images can be degraded by three major sources: atmospheric perturbation, imaging systems, and instrumental noises. The atmosphere can have several major impacts on remote sensing data (particularly the ones captured by passive sensors) such as absorption, scattering, and reflection of the solar radiation in the atmosphere. Imaging systems induce artifacts and noise such as speckles in synthetic aperture radars (SARs) and striping in hyperspectral images (HSIs) push-broom imaging systems. Instrumental (sensor) noise includes thermal (Johnson) noise, quantization noise, and shot noise (for optical imagery). To compensate for the atmospheric effects, atmospheric corrections should be applied. The noises and artifacts induced by imaging systems and the instruments often are treated by image processing and machine learning techniques. Generally speaking, the observed degraded remote sensing image can be modeled with
\begin{equation}\label{eq: RSmodel}
	{\bf H} = {\bf M}\times{\bf X} + {\bf S}+{\bf N},
\end{equation}
where ${\bf H}$ represents the observed image, ${\bf M}$ denotes the multiplicative noise (also denotes the blur kernel in deblurring problems), ${\bf S}$ is the independent additive sparse noise and is often assumed to have a Laplace distribution, and ${\bf N}$ is an additive noise (which may be modeled as signal-independent when large numbers of photons are collected, or with a signal-dependent variance in low-flux regime). The multiplication $\times$ depending on the application might be element-wise or matrix multiplication e.g., for despeckling and deblurring, respectively.
Therefore, the image restoration task considered in this paper is to estimate the unknown image ${\bf X}$ via the observation ${\bf H}$. Generally, inverse  image restoration tasks such as denoising or deblurring can be formulated in the framework of an optimization problem given by
\begin{equation}\label{eq: ImRec}\textcolor{black}{
(\hat{\bf X}, \hat{\bf S}, \hat{\bf M})=\arg\min_{{\bf X,S,M}}Q({\bf H},{\bf X}, {\bf S}, {\bf M})+\lambda R({\bf X, S, M}),}
\end{equation}
where the function $Q$ defines the fidelity with respect to the observed data, $R$ is a regularizer (or penalty) function selected according to the prior knowledge, and $\lambda$ denotes the tuning parameter which
balances 
the fidelity and the penalty terms. We should add that problem (\ref{eq: ImRec}) can be defined subject to equality and inequality constraints.

Multiplicative noise is typical in coherent imaging systems such as SAR or ultrasound imaging. Depending on their relative phase, \textcolor{black}{all} elementary scattered signals may add constructively or destructively, resulting in bright and dark spots in the image. Sparse noises such as salt and pepper noise, missing pixels, missing lines, and other outliers usually appear in the observed image mostly due to malfunctioning imaging instruments. In optical imaging systems, shot noise or photon noise is modeled as a signal-dependent additive noise whose variance depends on the signal level. Other instrumental noises such as thermal noise and quantization noise are modeled by signal-independent Gaussian additive noise \cite{John, Landgrebe}. 

Another common type of noise that sometimes degrades remote sensing images is called "pattern noise" and is generally due to the imaging system. For instance, the most common one, described in detail in this paper, is called striping. It is caused by push-broom imaging systems where the target scene is scanned line by line and the image lines are acquired at different wavelengths by an area-array detector. Striping noise is often due to calibration errors and sensitivity variations of the detector \cite{MODIS_Destrip, Patternoise}. For simplicity reasons, the striping noise is often modeled as a sparse independent additive noise, however, it depends on the signal level and the position of detectors of the acquisition array in the cross-track direction \cite{BR_HS_DN_Rev,Acito_Destrip}. 

\begin{figure}[htbp]
    \centering
    \includegraphics[width=0.48\textwidth]{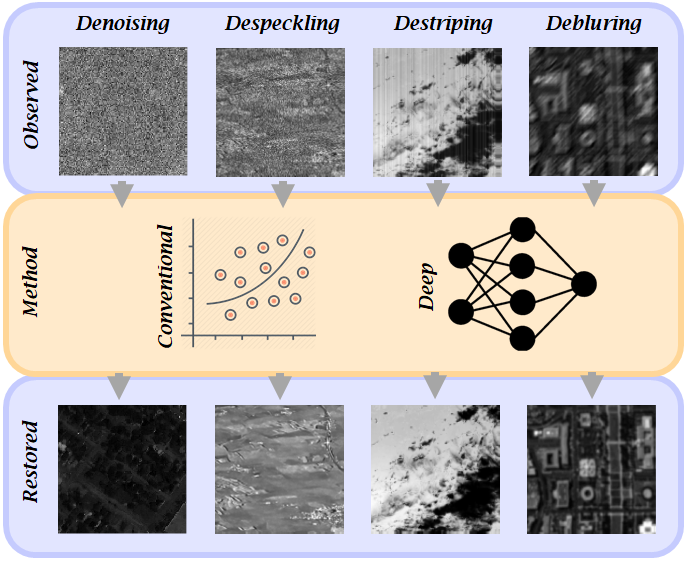}
    \caption{Four remote sensing restoration applications studied in this paper. Denoising, despeckling, destriping, and deblurring are considered as major trends in remote sensing data restoration. The restoration methodology are categorized in two groups i.e., deep learning-based methods and conventional ones.}
    \label{fig:GAbs-intro}
\end{figure}

By taking the aforementioned points into account, in this paper, we provide overviews for remote sensing image restoration which fall into the following topics: denoising, despeckling, destriping, and deblurring \textcolor{black}{(see Fig. \ref{fig:GAbs-intro})}. It should be noted that the field of image restoration is much broader than those topics. However, the other topics either are not yet major trends in remote sensing (e.g., dehazing) or are subject to considerable attention (e.g., resolution enhancement), which would need to be addressed in a separate paper. 

\subsection{Statistics}
\begin{figure}[htbp]
\begin{center}
    \includegraphics[width=0.5
    \textwidth]{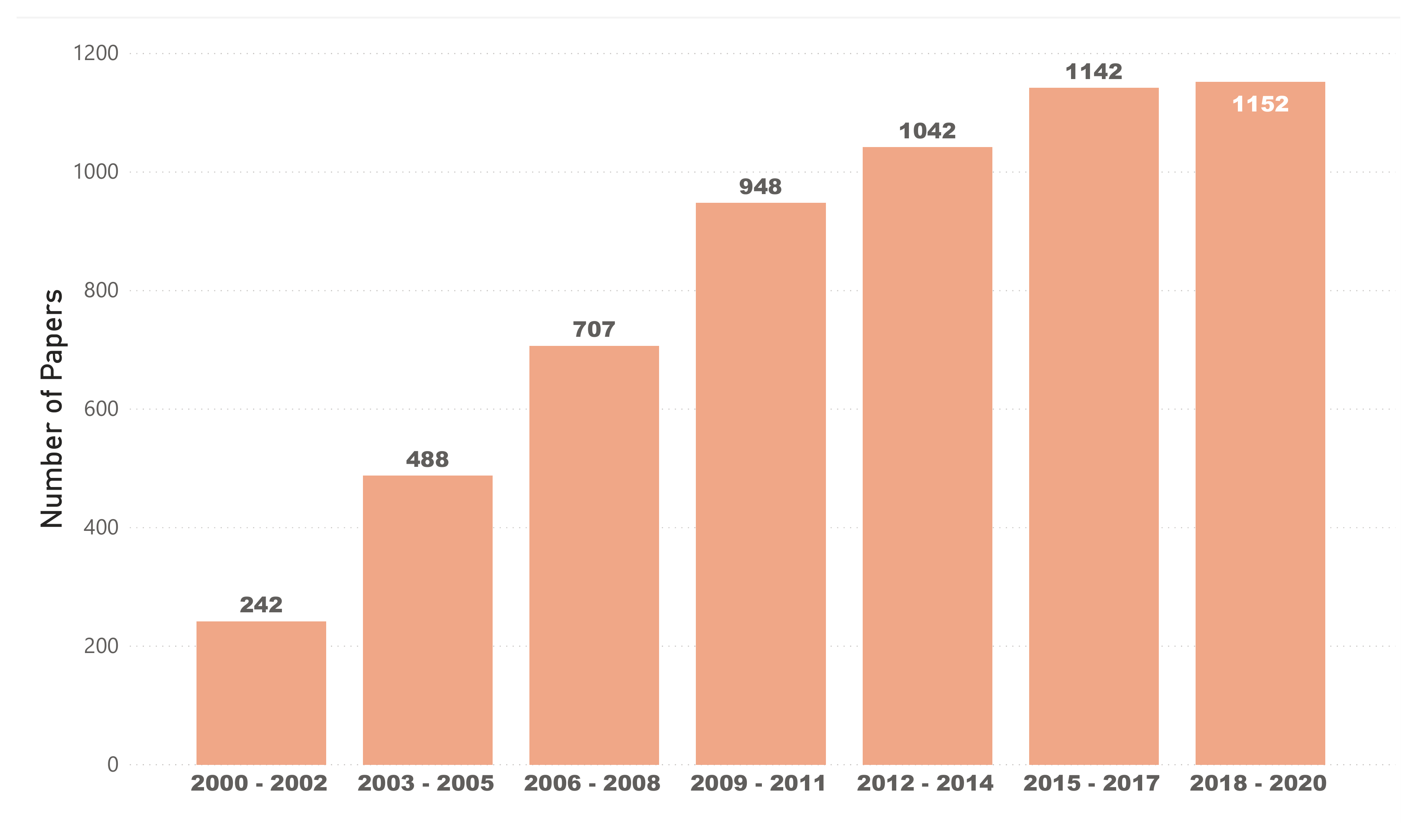}
 \end{center}
   \caption{The number of papers published by IEEE on the subject of remote sensing data restoration.}
\label{fig:stats}
\end{figure}

Fig. \ref{fig:stats} shows the dynamics of the important subject of denoising and restoration in the remote sensing community. The reported numbers include magazines, scientific journals, and conference papers published by IEEE on this particular subject using “remote sensing” and "(denoising, restoration, or noise reduction)" as the main keywords used in the abstracts. In order to highlight the number of papers, the time period has been split into a number of equal time slots (i.e., 2000–2002, 2003–2005, 2006–2008,2009–2011, 2012–2014, 2015–2017, and 2018–2020 (December 31st)). \textcolor{black}{The} steadily increasing number of papers reveals the popularity of this subject.

\subsection{Contribution}
In this paper, we provide a unique overview of the state-of-the-art restoration techniques proposed for two major remote sensing imaging systems i.e., SAR and hyperspectral. The restoration approaches considered in this paper include major paradigms in noise and artifact removal in the literature of SAR and HSI, i.e., denoising, despeckling, destriping, and deblurring techniques. The similarity of those applications motivated us to provide this overview to favor idea cross-fertilization between these two domains. A systematic study is carried out on the literature to reveal the evolution from the conventional techniques to deep learning techniques over the past decades. \textcolor{black}{We provide a systematic literature review by categorizing the most established methods proposed in the literature. Then, we select representative methods depending on the application and the multiplicity of the open-source codes from each category for the experimental results. The representative methods are not necessarily state-of-the-art. Please note that they belong to different periods throughout the evolution of the applications. However, they are well-established approaches and were effective at the time.} The uniqueness offered by this paper can be attributed to two different factors. First, the comprehensive overview is given in a way that draws a comparison of the state-of-the-art deep learning techniques with respect to the conventional techniques. This is greatly valuable due to the rapidly growing deep learning field and its influence on the remote sensing application. More importantly, the experiments carried out in this review paper reveal the advantages of deep learning techniques compared with the conventional techniques. Second, this paper is accompanied by a unique collection 
of libraries for conventional and deep learning restoration techniques which will be greatly beneficial to the remote sensing community particularly the young and new researchers in the field. As a result, this work can fast-forward the community to grow in the field.
\subsection{Notation}
\label{sec: Notation}
In this paper, the number of bands and pixels in each band are denoted by $p$ and $n=(n_1 \times n_2)$, respectively. Matrices are denoted by bold and capital letters, column vectors by bold letters, the element placed in the $i$th row and $j$th column of matrix ${\bf X}$ by $x_{ij}$, the $j$th row by ${\bf x}^T_j$, and the $i$th column by ${\bf x}_{(i)}$. The identity matrix of size $p\times p$ is denoted by ${\bf I}_p$. $\hat{\bf X}$ stands for the estimate of the variable ${\bf X}$. The Frobenius norm and the Kronecker product are denoted by $\left\|.\right\|_F$ and $\otimes$, respectively. Operator $vec$ vectorizes a matrix and $vec^{-1}$ is the corresponding inverse operator. $\mbox{tr}({\bf X})$ denotes the trace of matrix ${\bf X}$. 

\section{Restoration of Synthetic Aperture Radar}

SAR is a \emph{coherent} imaging technique. The coherent acquisition of the radar signal is essential: echoes collected at different locations along the track of the radar antenna can be numerically combined to refocus the wave field diffracted by the scene, the phase shift between SAR echoes captured under slightly different angles or dates reveals 3D locations and displacements in SAR interferometry. The downside of coherent imaging techniques is the \emph{speckle phenomenon}: measurements correspond to the coherent summation of several elementary responses, these responses may either add constructively (leading to a large echo) or destructively (leading to a very low echo). The outcome of this coherent summation is intimately related to the geometrical configuration of elementary scatterers and is modeled, for surfaces that are rough at the scale of the radar wavelength, by the Goodman's model \cite{goodman2007speckle}: the intensity of the SAR echo $\intensity$ at a given pixel is pertubed by a randomly fluctuating variable $\speckle$,
\begin{equation}
    \intensity =  \speckle \times\reflectivity\,,
\end{equation}
where the radar reflectivity is $\reflectivity$. In this multiplicative noise model, $\speckle$ is distributed according to a gamma law:
\begin{equation}
    \text{p}(\speckle) = \frac{L^L}{\Gamma(L)}\speckle^{L-1}\exp(-L\speckle)\,,
    \label{eq:speckle}
\end{equation}
where $\Gamma(\cdot)$ is the gamma function and $L$ denotes the number of looks, i.e., a parameter that accounts for the possible incoherent averaging of radar intensities during a pre-processing step ($L=1$ in the absence of such pre-processing).
Variable $\speckle$ has unitary mean and $\text{Var}[\speckle]=1/L$, leading to $\mathbb{E}[\intensity]=\reflectivity$ and $\text{Var}[\intensity]=\reflectivity^2/L$, which indicates that the intensity of SAR echoes suffers from fluctuations that are stronger in bright areas (large values of $\reflectivity$) than in dark areas (low values of $\reflectivity$).
To make speckle fluctuations signal-independent, several SAR image restoration techniques apply a logarithmic transform to SAR intensities.
The log-intensity $\logintensity$ is related to the log-reflectivity $\logreflectivity$ through:
\begin{equation}
    \logintensity = \logspeckle+\logreflectivity \,,
\end{equation}
where the log-speckle $\logspeckle$ follows a Fisher-Tippett distribution: 
\begin{equation}
    \text{p}(\logspeckle) = \frac{L^L}{\Gamma(L)}e^{L\logspeckle}\cdot \exp(-Le^{\logspeckle}) \,,
    \label{eq:logspeckle}
\end{equation}
and the variance is constant: $\text{Var}[\logspeckle] = \psi(1,L)$ ($\psi(1, L)$ is the polygamma function of order $L$, see e.g. \cite{abramowitz1965handbook}). Averaging speckle samples in log domain requires an adequate debiasing step since the log-speckle has a non-zero mean: $\mathbb{E} [\logspeckle] = \psi(L)-\log(L)$ ($\psi$ is the digamma function).

Beyond the intensity of the back-scattered echoes, SAR images can also capture information of the phase and polarization of the wave. Such additional information is central in several applications of SAR imaging based on polarimetric analysis and classification, interferometry, differential interferometry and tomography. At each pixel, a diffusion vector $\diffvector$ is collected ($\diffvector\in\mathbb{C}^2$ in SAR interferometry (InSAR), $\diffvector\in\mathbb{C}^3$ in SAR polarimetry (PolSAR), $\diffvector\in\mathbb{C}^6$ in PolInSAR, $\diffvector\in\mathbb{C}^N$ with $N\geq 3$ in SAR tomography). Due to the speckle phenomenon, $\diffvector\in\mathbb{C}^K$ fluctuates according to a complex circular Gaussian distribution:
\begin{equation}
    \text{p}(\diffvector|\covmat) = \frac{1}{\pi^K\,|\covmat|}\exp(-\diffvector\hermtranspose \covmat^{-1} \diffvector)\,,
        \label{eq:diff}
\end{equation}
where the covariance matrix $\covmat\in\mathbb{C}^{K\times K}$ characterizes the image surface (i.e., it contains the interferometric and/or polarimetric information), $\diffvector\hermtranspose$ is the Hermitian transpose of column vector $\diffvector$, and $|\covmat|$ is the determinant of the covariance matrix $\covmat$. To access that information, the sample covariance $\empcov$ is typically computed by averaging over a small local window of L pixel:
\begin{equation}
    \empcov = \frac{1}{L}\sum_{\ell=1}^L {\diffvector_\ell^{\phantom{\dagger}}} {\diffvector_\ell\hermtranspose}\,.
    \label{eq:empcovsar}
\end{equation}
Due to speckle, the sample covariance matrix is noisy: it fluctuates according to Wishart's distribution
\begin{equation}
    \text{p}(\empcov|\covmat,L) = \frac{L^{LK}|\empcov|^{L-K}}{\Gamma_K(L)|\covmat|^L}\exp(-L\mbox{tr}(\covmat^{-1}\empcov))\,.
        \label{eq:wishart}
\end{equation}
For large values of $L$, speckle fluctuations are limited but this corresponds to averaging many pixels which represents a dramatic resolution loss.

The statistical models of speckle given in equations (\ref{eq:speckle}), (\ref{eq:logspeckle}), (\ref{eq:diff}), and (\ref{eq:wishart}) are at the core of SAR restoration techniques.

SAR speckle reduction is a challenging problem for several reasons:
\begin{itemize}
    \item SAR images have a high-dynamic range, with contrasts that span several orders of magnitude between scattering surfaces with moderate roughness and the strong echoes produced by multiple reflections on man-made structures (dihedral and trihedral configurations of the ground and building walls, metal fences, power poles);
    \item due to the heavy-tailed distribution of speckle (eq. \ref{eq:speckle}), maximum a posteriori estimators require solving non-convex minimization problems;
    \item multi-variate SAR modalities like InSAR, PolSAR, PolInSAR, and SAR tomography involve the estimation of complex-valued covariance matrices.
\end{itemize}

\begin{figure*}[!t]
\includegraphics[width=\textwidth]{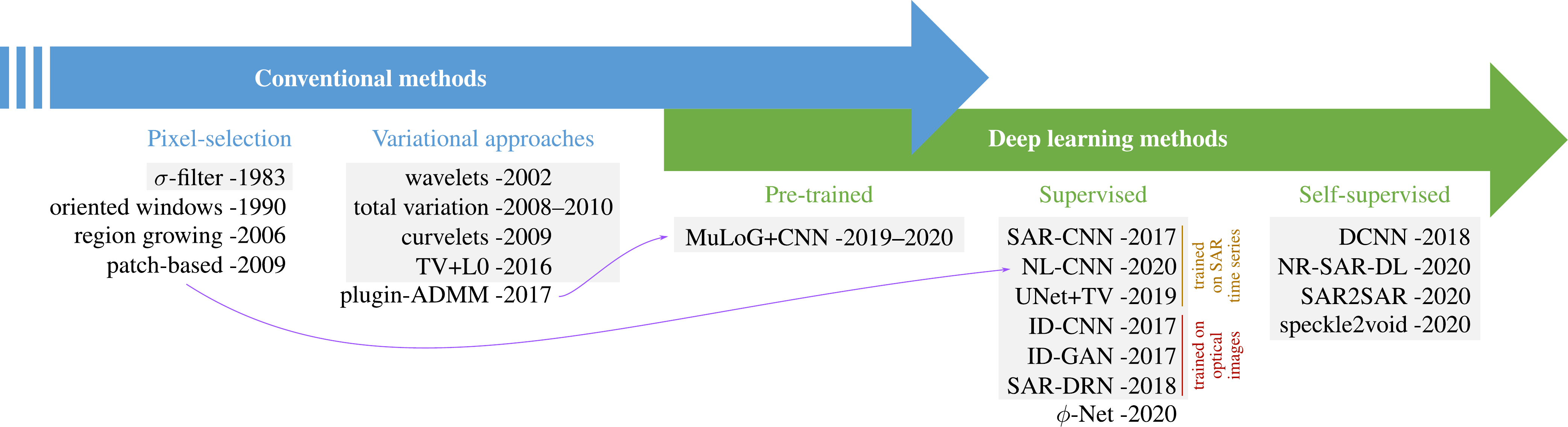}
\caption{Some representative methods for SAR image restoration, with a special emphasis on deep learning techniques. Methods in gray boxes are designed for SAR intensity images while the other methods can be applied to polarimetric and/or interferometric SAR images.}
\label{fig:summarySAR}
\end{figure*}

\textcolor{black}{Fig. \ref{fig:summarySAR} shows some representative methods for speckle reduction in SAR imaging, with a temporal evolution (symbolized by the colored arrows) from conventional methods which started emerging in the 1980s to the more recent deep learning methods.}

\subsection{Conventional Techniques}
The strong, signal-dependent, non-Gaussian fluctuations corrupting SAR intensity images have driven the development of numerous restoration approaches. The early approaches were based on local filters, i.e., on the averaging of intensities within a small window. Due to the strong speckle fluctuations, the use of small windows is not sufficient to reach a satisfying level of smoothing. Increasing the size of the averaging window, however, leads to an unacceptable resolution loss. Some mechanism is required to prevent the blurring of point-like structures and edges. Lee's $\sigma$ filter \cite{lee1983digital} prevents the mixing of large intensities and lower intensities by restricting the averaging to values belonging to a range centered on the value of the central pixel. The local selection of an oriented averaging window \cite{lopes1990adaptive} or the selection of pixels by region growing \cite{vasile2006intensity} help to reduce the blurring of edges. 
\textcolor{black}{
A robust approach to identify similar pixels consists of comparing image patches \cite{NLM,deledalle2011patch,deledalle2012compare}. \textit{So-called} \emph{non-local methods} perform a weighted averaging with weights derived from patch similarities \cite{deledalle2009iterative,parrilli2011nonlocal,cozzolino2013fast,deledalle2010nl,chen2010nonlocal,deledalle2014exploiting,torres2014speckle,deledalle2015nl}.
All these pixel-selection methods form an estimate $\hat{x_i}$ (or $\widehat{\covmat}_i$ in SAR polarimetry and SAR interferometry) at pixel $i$ based on the noisy observed values $h_j$ (or $\empcov_j$) in the neighborhood $\mathcal{N}_i$ centered on pixel $i$, using a weighted averaging:
\begin{align}
    \widehat{x_i} = \sum_{j\in\mathcal{N}_i} w_{i,j} h_j\quad\text{ and }\quad\widehat{\covmat}_i=\sum_{j\in\mathcal{N}_i} w_{i,j} \empcov_j\,,
\end{align}
where the strategy to derive the weights $w_{i,j}$ from the data varies according to the pixel-selection method.}

\textcolor{black}{
Rather than explicitly selecting similar pixels, variational approaches (second column of Fig.\ref{fig:summarySAR}) define the estimator as that achieving a trade-off between data fidelity and adequacy with the prior model, i.e., a regularization term, as stated in Eq.(\ref{eq: ImRec}). Edge-preserving priors such as the total variation (TV) \cite{aubert2008variational,denis2009sar,bioucas2010multiplicative}, image decomposition priors like TV+$\ell_1$ and TV+$\ell_0$ \cite{denis2010exact,lobry2016multitemporal}, or wavelet and curvelet transforms \cite{xie2002sar,durand2009multiplicative} have been considered for the restoration of intensity images. The data-fidelity term $Q$ in Eq.(\ref{eq: ImRec}) can be derived from the gamma distribution (Eq.(\ref{eq:speckle})) or the Fisher-Tippett distribution (Eq.(\ref{eq:logspeckle})). The latter has the advantage of being convex, and thus, easier to minimize. The regularization can either be applied to the reflectivities or to the log-transformed reflectivities, leading to minimization problems of the form}
\begin{align}
    \textcolor{black}{{\hat {{\bf x}}}=\arg\min_{{\bf x}}~L\sum_{i=1}^n\left( x_i+\frac{h_i}{x_i}\right)+\lambda\phi({ \bf x}),}
\end{align}
\textcolor{black}{or}
\begin{align}
    \textcolor{black}{{\hat {\tilde{\bf x}}}=\arg\min_{\tilde{\bf x}}~L\sum_{i=1}^n\left( \tilde{x}_i-\tilde{h}_i+\exp(\tilde{h}_i-\tilde{x}_i)\right)+\lambda\phi({ \tilde{\bf x}}),}
\end{align}
\textcolor{black}{where ${\bf x}\in\mathbb{R}_+^n$ is the $n$-pixels restored reflectivity image, ${\bf h}\in\mathbb{R}_+^n$ is the $n$-pixels noisy SAR intensity image, in linear scale, and $\tilde{\bf x}\in\mathbb{R}^n$ and $\tilde{\bf h}\in\mathbb{R}^n$ are their equivalent in log-scale.}

\textcolor{black}{Patch-based and variational methods can be combined, which is particularly beneficial in the context of interferometric phase estimation to reconstruct buildings \cite{ferraioli2017parisar}.
}




The specificities of speckle in SAR images have justified the design of dedicated restoration methods, often inspired by the continuing progress in the field of natural image denoising. Adapting those methods can, however, be a tedious task. A way to circumvent these adaptations is to tackle speckle reduction using a plugin ADMM approach, i.e., decomposing the image reconstruction process into an alternation of non-linear steps to account for speckle statistics and Gaussian denoising steps that can be performed by any off-the-shelf Gaussian denoiser \cite{deledalle2017mulog}.

\begin{table}[htbp]
    \centering\addtolength{\tabcolsep}{-3pt}
    
    \caption{ \textcolor{black}{Processing time (in seconds) of the despeckling techniques applied to an image patch of size $500\times500$ pixels. Experiments have been carried out on an Intel Xeon CPU at 3.40 GHz and an Nvidia K80 GPU. For Speckle2Void and SAR2SAR, the experiment has been performed on an Nvidia 1080 GPU.}}
    \begin{tabular}{cccccc}
		\toprule
    &SAR-BM3D & NL-SAR & MuLoG+CNN  & Speckle2Void & SAR2SAR \\ 	\midrule
    Time(s)&73.89 & 111.28 & 80.43 & 3.64 & 0.99 \\
		\bottomrule 
	\end{tabular}  
	\label{tab:time_SAR}

\end{table} 

\begin{figure*}[htbp]
\textcolor{black}{
\begin{center}
    \includegraphics[width=0.99\textwidth]{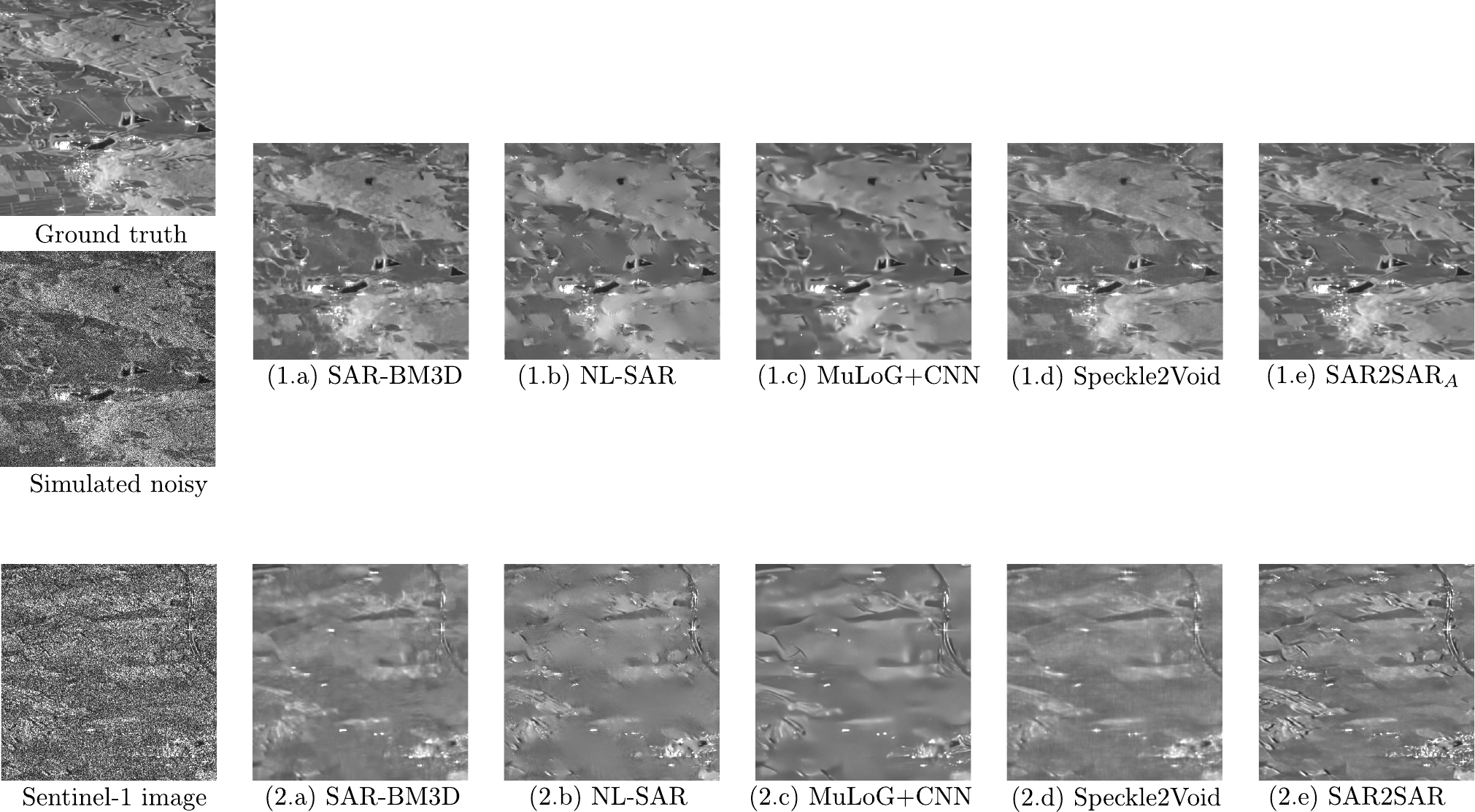}
 \end{center}
   \caption{Comparison of several speckle reduction methods for SAR imaging in a numerical simulation (top row) and on an actual single look Sentinel-1 image (bottom row). A ground truth image is obtained from a large temporal series of Sentinel-1 images and an image with simulated speckle is obtained; (1.a-1.e) restorations are produced by 2 conventional techniques and 3 deep-learning methods; The depicted images refer to a noisy instance of the "Limagne" dataset in Table \ref{table:comparison_psnr}. 
   (2.a-2.e) restoration results are obtained by the same algorithms on an actual single look Sentinel-1 image. Computational time for each method is given in Table \ref{tab:time_SAR}}. }
\label{SAR_results}
\end{figure*}

\begin{table*}[htbp]
\centering
\caption{Comparison of denoising quality in terms of PSNR on
amplitude images. For~each ground truth image, 20 noisy
instances are synthetically generated according to Goodman's speckle model, Eq.(\ref{eq:speckle}). 1$\sigma$ confidence intervals are
given. Per-method averages are given at the bottom.}
\begin{tabular}{l c c c c c c}
\toprule
Images  & Noisy & SAR-BM3D & NL-SAR & MuLoG+CNN & Speckle2Void & \textcolor{black}{SAR2SAR\textsubscript{A}}  \\
\midrule
Marais 1    & 10.05 $\pm$ 0.0141 & 23.56 $\pm$ 0.1335 & 21.71 $\pm$ 0.1258 &  23.39 $\pm$ 0.0608 & 24.89$\pm$0.1021 & \textcolor{black}{\textbf{25.73}$\pm$0.1251} \\
Limagne     & 10.87 $\pm$ 0.0469 & 21.47 $\pm$ 0.3087 & 20.25 $\pm$ 0.1958 &  21.16 $\pm$ 0.0249 &{23.40}$\pm$0.1206 & \textcolor{black}{\textbf{24.45}$\pm$0.1190}\\
Saclay      & 15.57 $\pm$ 0.1342 & 21.49 $\pm$ 0.3679 & 20.40 $\pm$ 0.2696 &  {21.88} $\pm$ 0.2195 & 19.00$\pm$0.4814 & \textcolor{black}{\textbf{23.60}$\pm$0.4368}\\
Lely        & 11.45 $\pm$ 0.0048 & 21.66 $\pm$ 0.4452 & 20.54 $\pm$ 0.3303 &  {22.17} $\pm$ 0.2702 & 19.28$\pm$0.5745 & \textcolor{black}{\textbf{23.67}$\pm$0.5415}\\
Rambouillet &  8.81 $\pm$ 0.0693 & {23.78} $\pm$ 0.1977 & 22.28 $\pm$ 0.1132 &  23.30 $\pm$ 0.1140 & 21.47$\pm$0.4879 & \textcolor{black}{\textbf{24.16}$\pm$0.3846}\\
Risoul      & 17.59 $\pm$ 0.0361 & 29.98 $\pm$ 0.2638 & 28.69 $\pm$ 0.2011 &  \textbf{30.85} $\pm$ 0.1844 & 21.42$\pm$0.1187 & \textcolor{black}{30.68$\pm$0.2302}\\
Marais 2    &  9.70 $\pm$ 0.0927 & 20.31 $\pm$ 0.7833 & 20.07 $\pm$ 0.7553 &  21.00 $\pm$ 0.4886 & {25.04}$\pm$0.2223 & \textcolor{black}{\textbf{26.63}$\pm$0.2154}\\
\midrule
Average    &  12.00 & 23.17 & 21.99 & 23.39 & 22.07 & \textcolor{black}{\textbf{25.56}} \\
\bottomrule
\end{tabular}
\label{table:comparison_psnr}
\end{table*}

\subsection{Deep learning-based Techniques }

Deep neural networks have the capability to learn application-specific patterns and adapt to non-Gaussian corruptions. Unsurprisingly, deep learning applications to the restoration of SAR images have flourished these last years\textcolor{black}{, see right part of Fig.\ref{fig:summarySAR}}. When designing a neural network, special care must be paid to the specificities of SAR imagery, in particular the high dynamic range of SAR intensities, the complex-valued definite-positive covariance matrices that characterize the interferometric or polarimetric information, the non-stationary noise variance, or the spatial correlations of speckle due to spectral apodization.

Training deep models requires a huge amount of data in order to generalize well. However, in SAR image denoising, there is a shortage of ground truth information to train supervised models, and \textit{ad-hoc} datasets have thus to be built to this aim. In this context, the typical signature of bright scatterers and the geometrical distortion of an acquisition system (such as layover, shadowing and foreshortening) require a careful application of canonical data augmentation techniques (e.g. rotation and mirroring). Alternatively, pre-trained models can be integrated into the MuLoG framework mentioned in the previous paragraph \cite{yang2019sar,dalsasso2020sar}.

\textcolor{black}{Among end-to-end supervised deep learning techniques, the pioneering work of Chierchia \textit{et al.} \cite{chierchia2017sar} 
is representative of training strategies that build ground-truth images from 
temporal stacks of co-registered SAR images.
Only the areas not affected by temporal changes are kept and the supervised training is performed using pairs of speckle-corrupted images and multi-temporal averages.
The network architecture, called \textit{SAR-CNN}, is inspired from the DnCNN \cite{zhang2017beyond}. 
Cozzolino \textit{et al.} \cite{cozzolino2020nonlocal} introduced a method to combine a non-local filtering based on patches and deep learning: \textit{NL-CNN} \cite{plotz2018neural}, using a similar supervised training strategy.
Building such a training set is far from easy: temporal changes are frequent and, if not properly accounted for, the network may end up producing a biased result. 
As an alternative, simulated speckle noise can be added to a ground-truth image obtained by temporal-averaging \cite{lattari2019deep}. Yet, due to speckle temporal correlations, this reference image may still present residual speckle fluctuations. To mitigate this phenomenon, a mild denoising step can be applied to the multi-looked image. Speckled images can then be produced by drawing random fluctuations according to the theoretical speckle distribution of Eq.(\ref{eq:speckle}), see for example \cite{dalsasso2020sar}. }

Alternatively, large datasets can be built by adding simulated speckle noise to natural images (see \textit{ID-CNN} \cite{wang2017sar}, \textit{ID-GAN} \cite{wang2017generative}, \textit{SAR-DRN} \cite{zhang2018learning}). The statistical distribution of natural images is, however, quite different from that of SAR images \textcolor{black}{(where point-like and linear structures are much more frequent)}, causing a problem referred to as \textit{domain gap}. \textcolor{black}{Moreover, methods developed under the assumption of i.i.d. speckle suffer from artifacts due to the spatial correlation of speckle when applied to actual SAR images \cite{dalsasso2020handle}. Combining a spatial loss with a spectral term \cite{vitale2019new} along with an edge-preserving term \cite{vitale2020edge} still produces artifacts in homogeneous areas. The oversampling and spectral windowing operations are indeed not taken into account by Goodman's speckle model. Recent advances in self-supervised approaches for AWGN suppression \cite{lehtinen2018noise2noise, krull2019noise2void, laine2019high} allow training deep models directly on noisy data. Adapting them to SAR images is, therefore, of utmost importance, given the inherent scarcity of speckle-free images.}

 \textcolor{black}{SAR time-series can be exploited to this aim \cite{boulch2018learning,ma2020sar,dalsasso2020sar2sar}. Self-supervised approaches that build upon the noise2noise framework \cite{lehtinen2018noise2noise} rely on the intuition that, given the random nature of noise (assuming a perfect temporal decorrelation), a model trained to predict, from a noisy acquisition, another noisy realization ends up predicting the common component: the underlying reflectivity. However, in practice, the two acquisitions must be sufficiently separated in time for temporal speckle decorrelation to occur. Changes are then also more likely to arise. A patch-similarity measure can be used to weight the loss function \cite{ma2020sar}. Alternatively, in \textit{SAR2SAR} \cite{dalsasso2020sar2sar} a compensation is applied to make sure that only the speckle component differs between the two dates. Training a network on actual SAR images makes it possible to learn the actual statistics of speckle and to capture the spatial correlations induced by the impulse response of the SAR imaging system. Networks trained with self-supervised strategies are readily applicable to actual SAR images: the reconstructed images do not suffer from artifacts due to the spatial correlations of speckle.}

\textcolor{black}{Instead of learning from multi-temporal SAR series, Molini \textit{et al.} \cite{molini2020speckle2void} proposed \textit{Speckle2Void}, an adaptation of the blind-spot CNN \cite{laine2019high} to single-look intensity SAR image despeckling, by extending the analysis carried out in \cite{molini2020towards}. This approach is at the crossroad between Bayesian modeling and deep learning. 
With the blind-spot CNN, the clean value of a pixel is obtained by combining the observed value at that pixel and an estimation based solely on the values of the neighbouring pixels. The network is trained to predict the central value of a window from the values of the neighbouring pixels. The quality of this estimation is evaluated by comparing the predicted value with the actual (noisy) value, which provides a reference-less way to train the network.
In this unsupervised algorithm, there is no need for a SAR time series and the training set can thus easily be created. However, it requires speckle to be spatially uncorrelated (so that the speckle realization at the central pixel be independent from the speckle realizations at neighboring pixels). 
To improve the robustness to residual speckle correlations, 
the network in \cite{molini2020speckle2void} is trained with a central spot of variable size, which prevents from relying too heavily on pixels in the immediate vicinity of the target pixel.} 

Adapting neural networks to complex data is challenging. In classical deep models, inputs are processed separately. However, the real and imaginary part of multi-channel complex SAR data are mutually correlated and they have to be jointly processed. To the best of our knowledge, there are no end-to-end deep learning-based techniques to suppress speckle from polarimetric SAR data. Sica \textit{et al.} \cite{sica2020phi} propose an adaptation of the U-Net to interferometric SAR data. In their \textit{$\phi$-Net} framework, the real part and the imaginary part of the complex interferogram are decorrelated and then fed as input to the network in two separate channels. Their denoised estimates, then, allow computing the estimated interferometric phase and coherence. Simulated data have been used for this aim. The adaptation of deep models to the restoration of the interferometric phase, however, is still at its early stages. We believe that this is a promising research direction and expect that progress will be made in the near future.

\textcolor{black}{Combining ideas from conventional methods such as patch similarity with deep neural networks is another direction in which we expect further developments, beyond the recent approaches proposed in \cite{denis2019patches,cozzolino2020nonlocal}.}

\subsection{Results and Comparisons}
\subsubsection{Quantitative Evaluations}
We first consider in the top row of Fig.\ref{SAR_results} the case of synthetic speckle added to a high-quality image (obtained by temporally averaging a long time-series of Sentinel 1 images \textcolor{black}{and by slightly filtering the multilooked image to suppress remaining speckle fluctuations and mitigate the effect of possible changes \cite{dalsasso2020sar}).} \textcolor{black}{Since synthetic speckle was generated starting from this high-quality image, quantitative restoration criteria like the peak signal to noise ratio (PSNR) can be computed.}
\textcolor{black}{Five despeckling methods are compared: two patch-based methods, SAR-BM3D \cite{parrilli2011nonlocal} and NL-SAR \cite{deledalle2015nl}, and three deep-learning methods, MuLoG+CNN \cite{deledalle2017mulog} (corresponding to the deep neural network DnCNN \cite{zhang2017beyond} applied iteratively within an ADMM loop) and the self-supervised algorithms Speckle2Void \cite{molini2020speckle2void} and SAR2SAR \cite{dalsasso2020sar} (by taking the network weights at the end of step A of the algorithm, conducted on simulated images).}
The images produced by each method provide a large improvement of the signal-to-noise ratio compared to the speckle-corrupted input image shown in \ref{SAR_results}(b): see Table \ref{table:comparison_psnr}. In terms of PSNR, deep-learning methods achieve to date the best results.

\subsubsection{Qualitative Evaluations}
Each method depicted in Fig.\ref{SAR_results} suffers from different artifacts: SAR-BM3D tends to create some periodic patterns, NL-SAR suppresses some fine structures and tends to oversmooth the texture of the forested areas (extended light grey areas in the image), MuLoG+CNN result is slightly blurry, and Speckle2Void introduces some distortions of the brightest echoes in urban areas. 

The application of despeckling methods to a real single-look Sentinel 1 image, in \textcolor{black}{the bottom row of} Fig.\ref{SAR_results}, tends to indicate that the deep-learning-based technique SAR2SAR provides qualitatively the most detailed restoration. Double-checking the presence of structures with the results of other methods seems a safe approach given the poor signal-to-noise ratio of the input image. Quantitative evaluation is often limited by the need for a ground truth (which limits the evaluation to simulated speckle) and the choice of metrics that fail to cover the various types of artifacts encountered in the range of available restoration methods. Qualitative evaluation thus remains an important step to assess the performance of a despeckling algorithm.

\textcolor{black}{Note that, to be fair, results are shown only on images that match the cases considered during network training.
The network Speckle2Void provided by the authors of 
\cite{molini2020speckle2void} has been trained on TerraSAR-X data after a speckle whitening step \cite{lapini2013blind}: while it can be directly applied to images corrupted by white noise (synthetic speckle case shown in Fig.\ref{SAR_results}(1.d)), speckle spatial correlations have been reduced on the actual SAR image of Fig.\ref{SAR_results}(2.d) thanks to a subsampling operation (though the mismatch of resolution between images acquired by TerraSAR-X and Sentinel-1 would probably require a fine-tuning of the network). Conversely, SAR2SAR has been trained in a three-step training strategy: the weights of step A, obtained by training on images corrupted by synthetic speckle noise, are used to produce the results of Fig\ref{SAR_results}(1.e), while the final weights, after fine-tuning the network on real Sentinel-1 SAR images \cite{dalsasso2020sar2sar} that display spatially-correlated speckle, are used to produce the result of Fig.\ref{SAR_results} (2.e)}


\subsubsection{Discussion}
\textcolor{black}{The remarkable ability of deep neural networks to preserve textured areas, sharp edges, linear and punctual structures explains their superior ability to restore intensity images compared to conventional techniques. Moreover, as can be seen from table \ref{tab:time_SAR}, these techniques are remarkably fast at test time. An important aspect in the field of speckle restoration is the possibility to handle the spatial correlations of speckle. These correlations have long been neglected in the statistical models used to derive estimators as well as in the synthetic speckle simulations used to produce training sets in supervised learning. A preprocessing step such as a spatial subsampling \cite{dalsasso2020handle} is then necessary before applying despeckling algorithms to actual SAR data. Most recent self-supervised deep-learning techniques tend to overcome this limitation.}

\textcolor{black}{Due to the increase of the dimension with polarimetric and/or interferometric SAR imaging, it becomes very challenging to train deep neural networks that cover the variability of spatial and polarimetric/interferometric patterns and that can handle complex-valued data. This is still an active area of research. In the mean time, conventional methods dominate for these types of data.}

\section{Hyperspectral Denoising }

\begin{figure}[htbp]
\centering
  \begin{tabular}{c}
   \includegraphics[width=.48\textwidth]{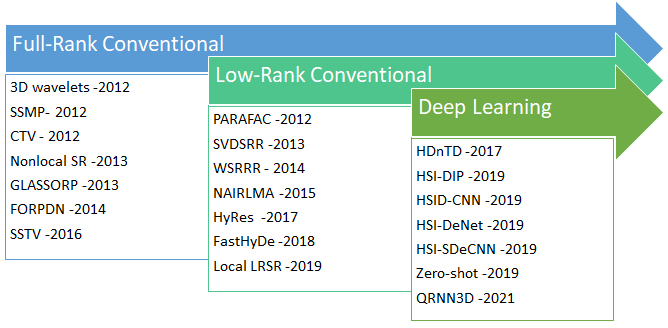}\\  \end{tabular}
  \caption{A glance of hyperspectral denoising literature over the past decade.}
\label{lit_GAbs}
\end{figure}

The observed HSI can be modeled as a combination of a true unknown signal and an additive noise. In general, an HSI ({\bf H}) can be modeled using a combination of vectors, 2D matrices, a 3D matrix, or a higher-order tensor. Using a matrix representation we have 
\begin{equation}\label{eq: HSImodel}
	{\bf H} = {\bf X} + {\bf N},
\end{equation}
where ${\bf H} \in \mathbb{R}^{n\times p}$ contains the observed spectral bands in its columns, ${\bf X} \in \mathbb{R}^{n\times p}$ is the noise free signal which needs to be estimated, and ${\bf N} \in \mathbb{R}^{n\times p}$ denotes the noise. The denoising task is to estimate the unknown noise-free signal ${\bf X}$. A conventional way is to use the penalized least-squares optimization:
\begin{equation}\label{eq: cost1}
	{\hat {{\bf X}}}=\arg\min_{{\bf X}}~\frac{1}{2}\left\| {\bf H}-{\bf X}\right\|^{2}_{F}+\lambda\phi({ \bf X}),
\end{equation}
where $\lambda$ determines the tradeoff between the fidelity term and the penalty term $\phi({ \bf X})$. We should note that the selection of the prior $\phi({ \bf X})$ plays an important role in the denoising performance. The noise is often assumed to be uncorrelated spectrally, i.e., ${\bf \Omega}=\mbox{diag}\left(\sigma^2_1,~\sigma^2_2,~\ldots,~\sigma^2_p\right)$ is the noise covariance matrix where $\sigma_i$ is the noise standard deviation of band $i$. Here, we review the conventional and deep learning-based denoising methods. Fig. \ref{lit_GAbs} glances throughout the hyperspectral denoising literature during the past decade. 
\subsection{Conventional Techniques }
   The conventional HSI denoising methods have evolved to capture both spatial and spectral correlation in different ways. The conventional approaches can be categorized into two independent groups: full-rank and low-rank approaches.
		\subsubsection{Full-rank Approaches}
		\label{subsub:3DModel}
		This group assumes full-rank for { \bf X} in the linear model (\ref{eq: HSImodel}). Conventional gray scale image denoising approaches such as wavelet denoising or mean filtering can be applied on HSIs band by band. In this way, the spectral information is ignored. On the other hand, hyperspectral data can be considered as a combination of spectral pixels and, then, signal denoising approaches can be used. However, the spatial information is, then, ignored in this way. For instance, the multiple linear regression (MLR) proposed in \cite{HySime} assumes that every band is a linear combination of the others. Hence, the $i$th band can be estimated by using the least-squares estimation. HSI denoising has been considerably improved by exploiting both spatial and spectral information. 3D modeling and filtering such as 3D wavelets \cite{RastiB} and 3D (blockwise) nonlocal sparse denoising (Nonlocal SR) methods \cite{Qianjstar} can be mentioned as the earliest attempts for spatial-spectral approaches. Later on, penalized least squares exploiting spatial \cite{Zelinski}, spectral \cite{FORPjour}, and spatial-spectral penalties \cite{Mixing,RastiSSSPIE} were proposed for HSI denoising. Those approaches proposed different prior $\phi({ \bf X})$ to capture spatial and/or spectral correlations. For instance, in \cite{FORPjour} the first-order spectral roughness penalty (FORPDN) was suggested in the wavelet domain to capture the spectral correlations:
		\begin{equation}\label{eq: cost7}
	{\hat {\bf W}}=\arg\min_{\bf {W}}~\frac{1}{2}\left\|\left({\bf H-D}_2{\bf W}\right){\bf \Omega}^{-1/2}\right\|^{2}_{F}+\frac{1}{2}\sum_{l=1}^L\lambda^l\sum_{j=1}^p\left\|{\bf R}_p{\bf w}_j^l\right\|^2_{2},
\end{equation}
where ${\bf D}_2$ contains 2D wavelet bases, and ${\bf W}$ denotes the 2D wavelet coefficients. Note that ${\bf R}_p$ is a $(p - 1) \times p$ difference matrix. 
		In \cite{RastiSSSPIE}, a combination of the first-order spectral roughness penalty with a group $\ell_2$ as a spatial penalty (GLASSORP) was proposed. A spatial-spectral mixing prior (SSMP) \textcolor{black}{was} proposed for hyperspectral denoising in \cite{Mixing}.
		An efficient edge-preserving denoising approach is obtained with a prior $\phi({ \bf X})$ in (\ref{eq: cost1}) corresponding to the TV of the signal \cite{Rudin}. 
		\begin{equation}\label{eq: costtv}
	{\hat {{\bf X}}}=\arg\min_{{\bf X}}~\frac{1}{2}\left\| {\bf H}-{\bf X}\right\|^{2}_{F}+\lambda \left\|{ \bf X} \right\|_{TV},
\end{equation}
where $\left\|{\bf X}\right\|_{TV}$ is the isotropic TV quasi-norm of the matrix ${\bf X}$. In the case of hyperspectral images, ${\bf X}$ contains both spatial and spectral information and therefore the conventional isotropic TV quasi-norm can not effectively capture both spectral and spatial correlation. Therefore, several TV denoising approaches were developed for HSI denoising. Cubic TV (CTV) proposed in \cite{Yuan, ZhangH} was adapted to incorporate the spectral variations for HSI denoising. Those techniques assign different weights to the spatial variations compared to the spectral variations. For instance, CTV utilizes 
		\begin{equation}\label{eq: costCTV}
	{\hat {{\bf X}}}=\arg\min_{{\bf X}}~\frac{1}{2}\left\| {\bf H}-{\bf X}\right\|^{2}_{F}+\lambda\left\|\sqrt{({\bf D}_v{\bf X})^2+({\bf D}_h{\bf X})^2+\beta ({\bf X}{\bf R}_p^T)^2}\right\|_1,
\end{equation}	
where the trade-off between spectral and spatial variations is controlled using parameter $\beta$. ${\bf D}_h$ and ${\bf D}_v$ are the matrix operators for calculating the first-order vertical and horizontal differences of a vectorized image, respectively. Spatio-spectral TV was proposed in \cite{SSTV} which exploits anisotropic spatial-spectral TV penalties for hyperspectral denoising. 
		\subsubsection{Low-Rank Approaches}
		\label{subsub: LR}
		Due to the high spectral dimension of HSI, low-rank modeling has been found more efficient for HSI denoising \cite{RastiPhDThesis}. In the additive noise model, it is assumed that ${\bf X}$ is low-rank. The low-rank property can be applied in two ways: (1) via a low-rank model 
		\begin{equation}\label{eq: HSImodelLR}
	{\bf H} = {\bf F}{\bf V}^T + {\bf N},
\end{equation}
where ${\bf F}$ and ${\bf V}$ are of rank $r$ (the rank $r$ is typically much smaller than both the number of bands $p$ and the number of pixels per band $n$) or (2) via the optimization problem using a low-rank penalty (or constraint) such as the nuclear norm: 	
\begin{equation}\label{eq: costlrtv}
	{\hat {{\bf X}}}=\arg\min_{{\bf X}}~\frac{1}{2}\left\| {\bf H}-{\bf X}\right\|^{2}_{F}+\lambda_1\phi({ \bf X})+\lambda_2 \left\|{ \bf X} \right\|_*,
\end{equation}
where $\left\|{\bf X}\right\|_*$ is the nuclear-norm of the matrix ${\bf X}$, obtained by ${\sum_{i} \sigma_i({\bf X})}$, i.e., the sum of the singular values. 

		Several hyperspectral denoising approaches were proposed using Tucker3 decomposition \cite{Lathauwer,Renard,KaramiJ,Letexier}. Parallel Factor Analysis (PARAFAC) \cite{PARAFAC} is another low-rank HSI denoising approach. In \cite{SVDSRconf}, a wavelet-based low-rank model and $\ell_1$ regularization was proposed for HSI denoising (SVDSRR). An automatic hyperspectral restoration technique (HyRes) was proposed in \cite{HyRes} using the $\ell_1$ penalized least squares and a low-rank wavelet-based model given by
					\begin{equation}\label{eq: cost122}
	{\hat {{\bf W}}}=\arg\min_{{\bf W}}~\frac{1}{2}\left\| {\bf H}-{\bf D}_2{\bf W}{\bf V}^T\right\|^{2}_{F}+\sum_{i=1}^r\lambda_i\left\|{\bf w}_{(i)}\right\|_1,
\end{equation}
		where ${\bf V}$ contains the spectral eigenvectors of ${\bf H}$ and is given by the singular value decomposition (SVD): $\textnormal{SVD}({\bf H})=\tilde{\bf U}\tilde{\bf S}{\bf V}^T$ and the hyperspectral Stein's unbiased risk estimator (HySURE) \cite{HySURE} used to select the model parameters. In \cite{WSRRR}, the wavelet-based reduced-rank (WSRRR) model was proposed for simultaneous HSI denoising and feature extraction using a non-convex optimization problem given by
\begin{align}\label{eq: WBPCA}
	(\hat{\bf V},\hat{\bf W})&=\arg\min _{{\bf W},{\bf V}}\frac{1}{2}\left\|{\bf H}-{\bf D}_2{\bf W}{\bf V}^T\right\|^{2}_{F}+\sum_{i=1}^r\lambda_i\left\|{\bf w}_{(i)}\right\|_1\\\mbox{s.t.}&~ {\bf V}^T{\bf V}={\bf I}_r,\nonumber
\end{align}	
which estimates the wavelet coefficients of the unknown signal and orthogonal subspace simultaneously. Low-rank TV regularization was also proposed in \cite{IGARSS14TV,RastiPhDThesis} for both HSI denoising and feature extraction:
\begin{equation}\label{eq: costLRTV}
	{\hat {{\bf F}}}=\arg\min_{{\bf F}}~\frac{1}{2}\left\| {\bf H}-{\bf FV}^T\right\|^{2}_{F}+\sum_{i=1}^r\lambda_i\left\|vec^{-1}({\bf f}_{(i)})\right\|_{TV}\,
\end{equation}	
where the notation $\|vec^{-1}({\bf f}_{(i)})\|_{TV}$ corresponds to the value of the TV (i.e., the sum of the magnitude of the spatial gradient over the whole image support), computed on the $i$-th column of matrix $\bf F$ after reshaping this column-vector as a 2D image. Noise-adjusted image recovery using low-rank matrix approximation (NAIRLMA) was suggested in \cite{NAILRMA} exploiting both low-rank and sparsity norms. Fast hyperspectral denoising (FastHyDe) proposed in \cite{FastHyDe} is also a low-rank technique that first projects the HSI into a subspace selected by HySime \cite{HySime} and, then, applies BM3D denoising \cite{BM3Dwiener} on the eigen images. The low-rank data are projected back into the original space after denoising. In \cite{8737679}, a local low-rank and sparse representation (Local LRSR) was suggested based on a weighted nuclear norm for HSI denoising in the presence of Gaussian noise. 

Spectral linear unmixing techniques are also considered as a low-rank HSI denoiser \cite{BR_UnDN}. However, denoisers based on unmixing are often vulnerable to endmember estimations. In other words, either the endmembers might not be estimated correctly or the noise power might affect the endmember estimation, leading, in both cases, to poor results. Therefore, some techniques extract endmembers from class averages defined by ground truth information \cite{DCerra_2014} or from a library of endmembers \cite{AErturk_2016}. Some methods were developed for performing the denoising and unmixing in a unified framework for boosting the performance of each other \cite{TInce_2019, JYang_2016}. Recently, Block-Gaussian-Mixture Priors have been proposed for both hyperspectral denoising and inpainting \cite{Teodoro}. The prior was applied on 3D patches from a subspace and then they were estimated using the minimum mean squared error (MMSE).
\subsection{Deep learning-based Techniques }
Deep learning-based denoising techniques are considered state-of-the-art in the signal and image processing community. DL-based denoising techniques designed for RGB images can often be used for HSI denoising, however, they cannot take into account the HSI-specific characteristics such as low-rank and spectral dependency. Since 2017, DL-based denoising techniques have emerged for HSI denoising. There is, however, one fundamental challenge for the use of DL-based methods for HSI denoising: the high-dimensionality of HSIs, which leads to a massive number of trainable parameters, and the absence of a comprehensive training database. The imbalance between the number of training samples and trainable parameters makes the network training cumbersome to achieve a universal DL-based denoising technique for HSIs.


The common feature of all the DL denoising techniques is to incorporate a convolutional block spatially to capture the spatial dependency. From the viewpoint of training, HSI DL-based denoising techniques can be divided into two groups: 
\subsubsection{Unsupervised/self-supervised}
Unsupervised techniques do not use any training set and only rely on the observed image and iteratively train the network. One example of such methods is deep HSI prior (HSI-DIP) proposed in \cite{HSI_DIP}, which is an unsupervised denoising technique. It is based on \textit{deep image prior} (DIP) \cite{DIP} which utilizes a convolutional encoder-decoder network to implicitly induce a universal regularizer (the so-called image prior) in inverse problems including denoising. In \cite{HSI_DIP}, 2D convolution was extended to a 3D one, however, it was not as efficient as the 2D one. Self-supervised techniques also rely on the observed image for training while they create their own training sets from the observed noisy image to train the deep network. In \cite{Self_SIP}, a self-supervised (Zero-shot) denoising technique was proposed for HSIs in which CNN was trained based on the observed image, which is assumed to be the target, and an input image, which is obtained by adding noise to the observed image. Therefore, the network can be only trained using the observed image.

\subsubsection{Supervised}
Supervised techniques require training sets to train the network. Most of the DL-based denoising techniques are supervised whose performances are highly dependent on the existence of enough training data. In HSI denoising, the deep networks are trained in three ways: (1) using a dataset by collecting many patches \cite{SSDeepRCNN}, (2) using a database containing many real HSIs such as ICVL \cite{ICVL} developed in \cite{HSI-DeNet}, or (3) using a database by simulating HSIs based on RGB images using deep generative networks \cite{3D-UNet}. Compared to unsupervised techniques, the training part is computationally very expensive, particularly in the two latter cases. However, a well-trained network is found to be more accurate than an unsupervised network. In \cite{HDnTD}, a CNN with trainable nonlinearity functions (HDnTD) was suggested for HSI denoising. The difference between HDnTD and the conventional CNN is that the CNN used for HDnTD also learns the non-linearity function while training the network. A CNN with multi-channel 2D convolutional filters was proposed in \cite{HSI-DeNet} for HSI denoising (HSI-DeNet). GANs were also examined for HSI denoising in \cite{HSI-DeNet} (HSI-DeGAN) where it was shown that it is less performant than the CNN. A spatial–spectral deep residual CNN (HSID-CNN) was proposed in \cite{SSDeepRCNN} in which 2D and 3D convolutional filters were combined to capture spatial and spatial-spectral correlation in HSIs, respectively. 3D Quasi-Recurrent Neural Network (QRNN3D) \cite{QRNN3D} applies separated 3D convolutions to the input image to generate a candidate component and a neural forget gate which was assumed as a sequence along the spectral axis. The subcomponents were obtained using different activation functions. The obtained sequences were used to extract the hidden states using a quasi-recurrent pooling function. The next layer was formed by concatenating the hidden states. A 3D U-Net was suggested in \cite{3D-UNet} using 2D convolutional filtering in spatial direction and 1D convolutional filtering in the spectral direction. Recently, a single denoising CNN-based framework (HSI-SDeCNN) was proposed in \cite{SDeCNN} in which, first, the hyperspectral bands were extended along the spectral direction so that all spectral bands have centered by the same number of adjacent bands. Then, one band at a time was restored by applying the CNN on a downsampled HSI with as many adjacent bands as the original number of bands. The downsampling was performed to improve the efficiency of the algorithm.    

\subsection{Results and Comparisons}
\textcolor{black}{Here, we compare the results of different hyperspectral denoising algorithms including 3D Wavelet \cite{RastiB}, FORPDN \cite{FORPjour}, SSTV \cite{SSTV}, NAIRLMA \cite{NAILRMA}, HyRes \cite{HyRes}, HSI-DIP \cite{HSI_DIP}, FastHyDe \cite{FastHyDe}, HSI-SDeCNN \cite{SDeCNN}, and HSI-DeNet \cite{HSI-DeNet}. We should note that 3D wavelets, FORPDN, and HyRes are parameter-free techniques. For FastHyDe we set the dimension of the subspace to 10 and for the other methods, the parameters are selected as suggested in the corresponding publications. For HSI-DIP, as suggested in \cite{DIP}, the hyperparameters of the network are tuned to obtain the optimum performance. HSI-DeNet is applied on every 10 band since the depth filter cannot be more than 10. Additionally, the default model (i.e., the model given for the noise parameter $\sigma=20$) was used which gives the best results in the experiments. In the case of SDeCNN, we set the noise parameter $\sigma$ to 1 since it gives the best results for the simulated data while for the real dataset, we used the default value suggested by its authors, i.e., $\sigma=20$.}

\subsubsection{Quantitative Evaluations}
\begin{figure}[htbp]
\centering
  \begin{tabular}{cc}
   \includegraphics[width=.23\textwidth]{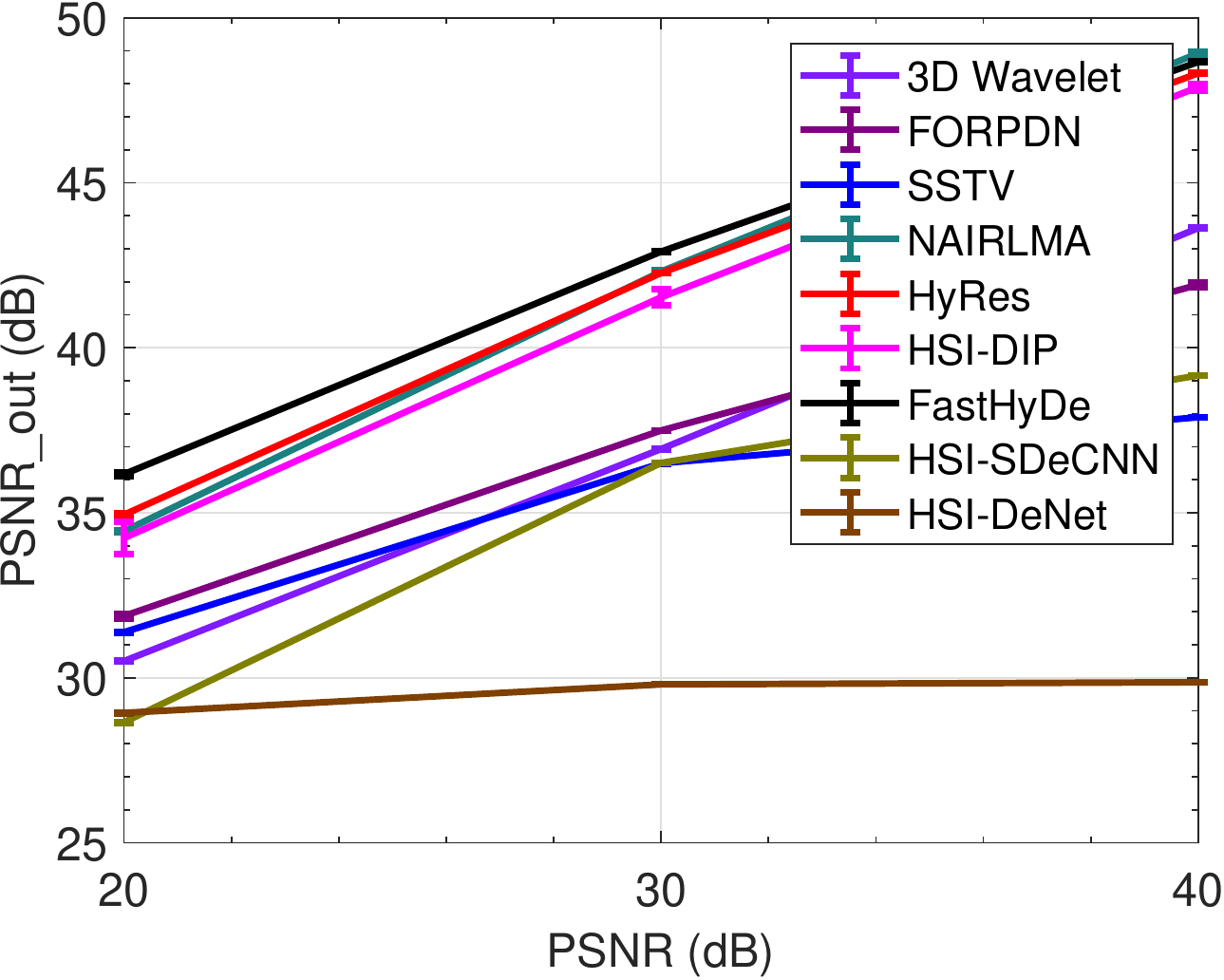}&
   \includegraphics[width=.23\textwidth]{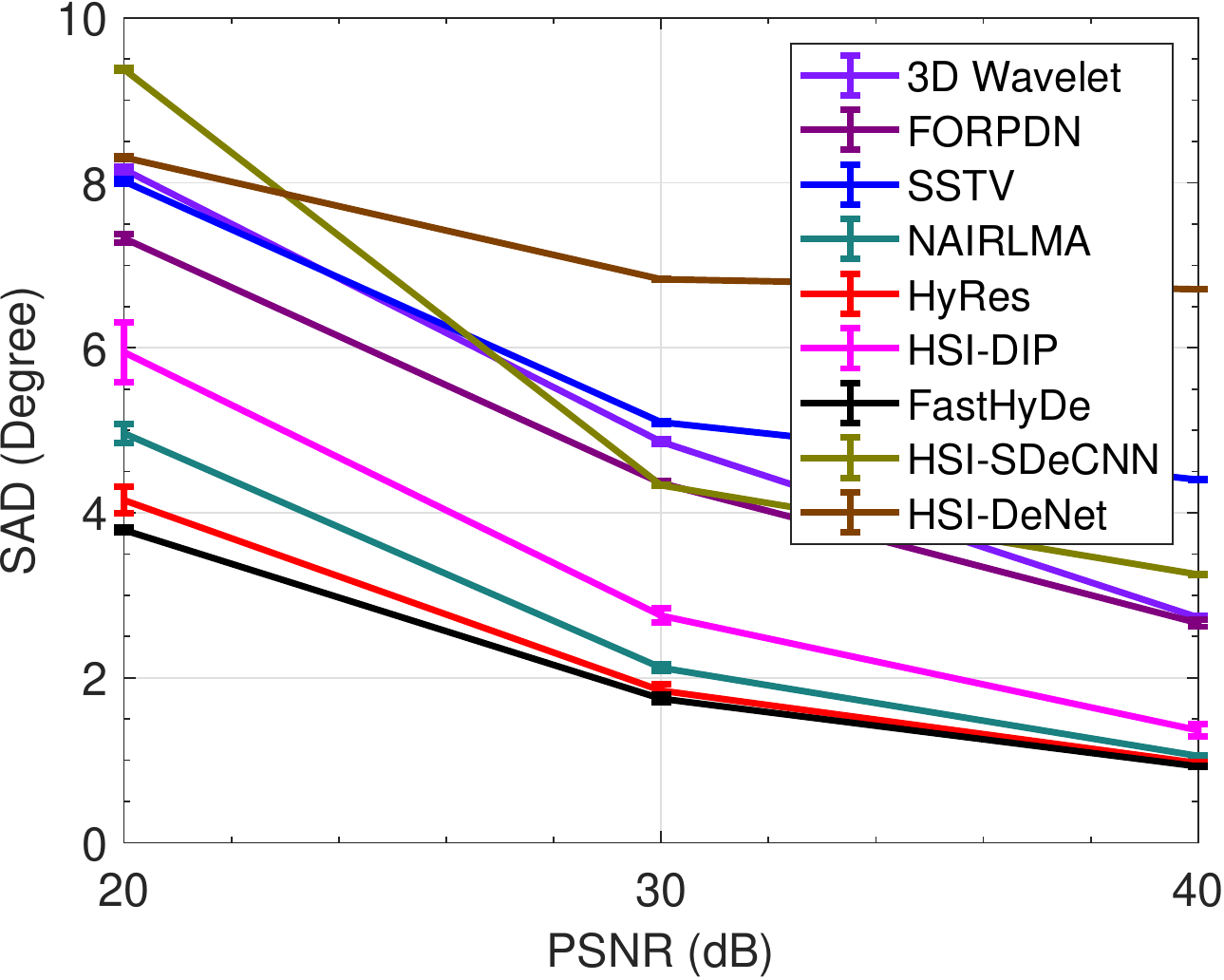}\\ 
  (a) PSNR & (b) SAD
  \end{tabular}
  \caption{\textcolor{black}{Results of denoising on simulated Washington DC Mall image for three different input noise levels (i.e., PSNR=20, 30, and 40 dB): (a) PSNR in dB (b) SAD in degree.}}
\label{SAD_PSNR}
\end{figure}

The hyperspectral denoising techniques are applied on a simulated noisy dataset and the results are compared for different levels of noise power, i.e., PSNR=20, 30, and 40 dB. \textcolor{black}{The uncorrelated (the same variance for all bands) zero-mean Gaussian noise was added to a portion of the Washington DC Mall dataset.} Fig. \ref{SAD_PSNR} (a) and (b) compare the results of the techniques in terms of PSNR and spectral angle distance (SAD). The results are mean values over five experiments and the standard deviations are shown using error bars. The results confirm that the low-rank techniques used in the experiments outperform the other techniques in terms of both PSNR and SAD. Additionally, HSI-DIP outperforms the full-rank conventional techniques, i.e., 3D wavelet, FORPDN, and SSTV. \textcolor{black}{On the other hand, the supervised DL-based techniques (i.e., HSI-SDeCNN and HSI-DeNet) perform poorly. }
\subsubsection{Qualitative Evaluations}
\begin{figure*}[htbp]
\centering
  \begin{tabular}{c}
   \includegraphics[width=.99\textwidth]{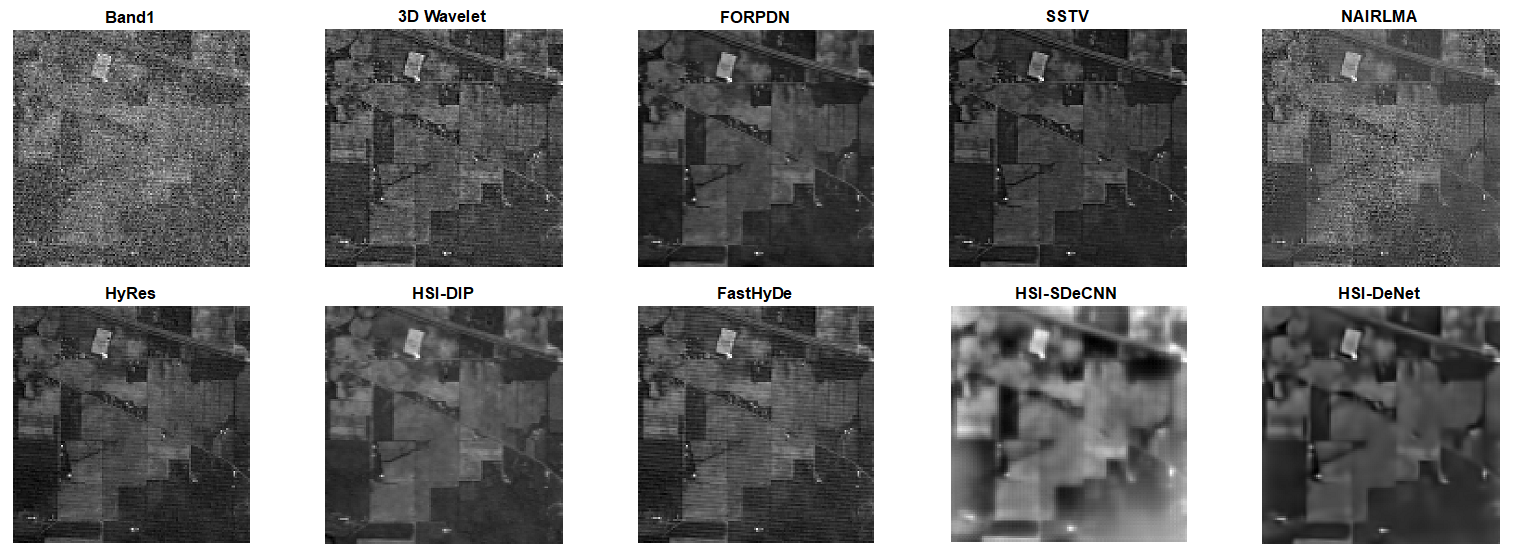}\\ 
  \end{tabular}
  \caption{\textcolor{black}{Results of denoising on Indian Pines. The first restored bands using different denoising techniques are visually compared with the observed noisy one.}}
\label{Real_DN}
\end{figure*}

For the real-image HSI denoising experiment, we apply all the denoising techniques on the Indian Pines dataset. The results of denoising for band 1 are compared visually in Fig. \ref{Real_DN}. The outcome of the visual comparison can be summarized as follows: NAIRLMA fails to restore band 1. \textcolor{black}{HSI-SDeCNN and HSI-DeNet show poor results and over-smoothed Band 1. We should note that HSI-SDeCNN was also applied to Indian Pines in \cite{SDeCNN}, however, the authors removed the noisy bands from the dataset  (resulting in a dataset having 206 bands) before applying the denoising techniques. Here, all 220 bands were used to evaluate the methods since the aim of HSI denoising techniques is to recover all the corrupted bands.} 3D Wavelet restores band 1 with moderate visual quality. SSTV, HyRes, and FastHyDe successfully and similarly reconstruct band 1. HSI-DIP also restores band 1 successfully, however, the restored band seems over-smoothed and blurred. FORPDN outperforms the other techniques visually. We should note that considering numerous bands existing in HSIs, this comparison might not reflect the performance of the techniques on the whole image. Band 1 is selected for comparison since it is noisy and it is often more challenging to recover the first and last few bands in HSI due to the absence of the adjacent bands.  \textcolor{black}{Processing times for the denoising techniques applied to the Indian Pines dataset are given in Table \ref{tab:timeDN}. All the denoising techniques were implemented in Matlab (2020b), except HSI-DIP which was implemented in Python (3.8). The reported processing times in this section were obtained using a computer with an Intel(R) Core(TM) i9-10980 HK processor (2.4 GHz), 32GB of memory, a 64-bit Operating System and an NVIDIA GEFORCE RTX (2080 Super) graphical processing unit. The reported results are mean values over five experiments. From the table, it can be observed that FastHyDe is the most efficient algorithm. We should note that the denoising algorithm BM3D utilized in FastHyDe was implemented in C. HyRes, FORPDN, and \textcolor{black}{HSI-SDeCNN} are also fast. Additionally, the results reveal that SSTV is computationally very expensive. It is worth mentioning that HSI-DIP \textcolor{black}{and HSI-DeNet} is computationally competitive compared with some of the conventional methods, which can be attributed to the efficiency of the GPU implementation.}

\begin{table*}[htbp]\centering\textcolor{black}{
	\caption{Processing time (in seconds) of the denoising techniques applied to the Indian Pines dataset.}
\begin{tabular}{cccccccccc}
	\toprule 
    &Wavelet3D & FORPDN & SSTV  & NAIRLMA & HSI-DIP & HyRes & FastHyDe &HSI-SDeCNN&HSI-DeNet\\ 	\midrule
    Time(s)&22.54 & 0.95 & 109.40 & 58.79 & 39.68 & 0.68 & 0.10 &2.17 &24.17\\
		\bottomrule
	\end{tabular}  }      
	\label{tab:timeDN}
\end{table*}
\subsubsection{Discussion}
The superiority of the low-rank techniques compared to the full-rank ones can be attributed to the spectral redundancy of HSIs. Low-rank techniques capture the most variations of the signal by projecting it into a subspace which helps to decorrelate the signal spectrally from the noise. Denoising techniques such as HyRes and FastHyDe further denoise the signal spatially in its subspace. HSI-DIP is an unsupervised denoising technique, i.e., the network is trained using observed data. Therefore, HSI-DIP does not outperform the well-established methods. \textcolor{black}{ HSI-SDeCNN exploits only one CNN model which is trained based on one dataset. Therefore, it cannot perform well for all types of HSI data with different noise levels. HSI-DeNet provides few models for few noise levels which, due to varying sensor and data characteristics, are not suitable for denoising HSIs. } We should note that, in the case of HSIs in particular HSI denoising, the training is often challenging due to the absence or the limited number of training sets. Therefore, Low-rank-based HSI denoising techniques still outperform the deep learning-based ones. Note that DL-based techniques outperform the conventional ones when there exist enough training data. However, in the case of HSIs in particular HSI denoising, the training is often challenging due to the absence or the limited number of training sets.

\section{Hyperspectral Destriping}

A push-broom-based hyperspectral imaging system would typically 
contain a thin slit, a diffraction grating, and a charge-coupled device (CCD) 2D array. The imaging system scans the target line by line 
and this line signal, then, is diffracted and separated into different wavelengths via the diffraction grating. Finally, the 2D CCD sensor records the diffracted wavelengths via the perpendicular axis and records the 1D spatial signal via the main axis \cite{gomez2008correction}. 

According to this imaging principle, the cause of the stripe noise can be mainly attributed to the failure of the CCD and slit. 
The non-uniform response of neighboring detectors due to the dark current generation, threshold variations, and gain or offset differences may unexpectedly generate a fixed-line pattern stripe noise \cite{Alessandro2001Analysis}. Additionally, instrument temperature variations result in dilation of the slit that changes its width and would also generate complex fixed-line pattern stripe noise. It is not possible to manufacture an ideal hardware that compensates for these fluctuations in HSIs.

\begin{figure*}[htbp]
\begin{center}
    \includegraphics[width=0.99\textwidth]{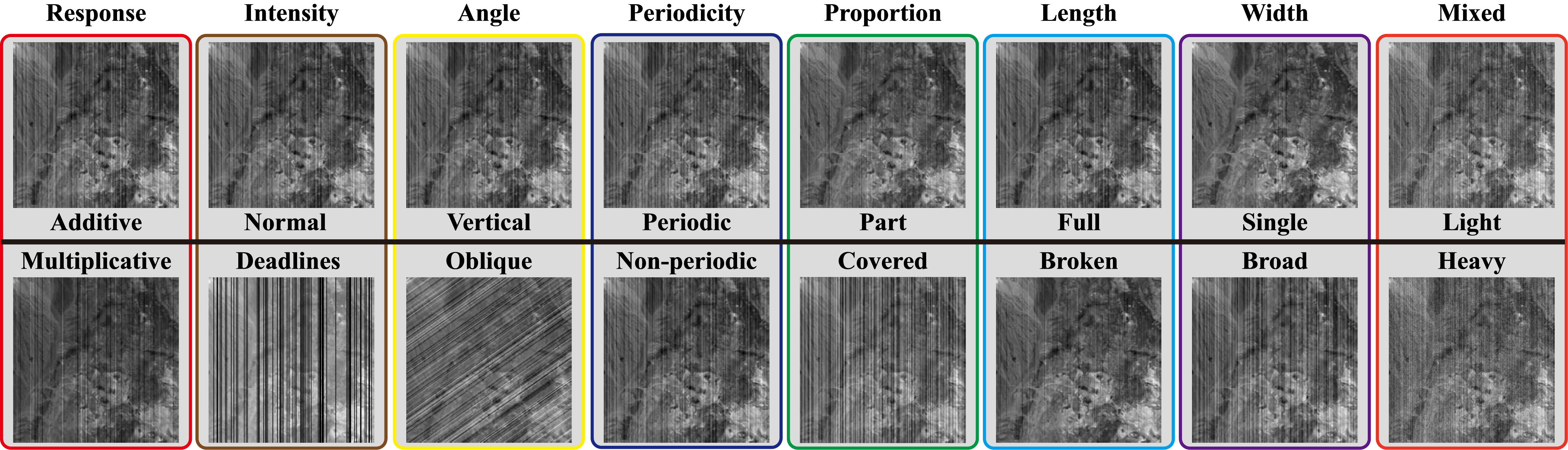}
 \end{center}
   \caption{Categories of stripe noise. We classify the stripes into eight categories according to their visual appearances, including (a) response, (b) intensity, (c) angle, (d) periodicity, (e) proportion, (f) length, (g) width, and (h) mixture of other noises.}
\label{Category}
\end{figure*}

\subsection{Categories of the HSIs Stripes}
Here, we classify the existing stripes into eight categories including response, intensity, angle, periodicity, proportion, length, width, and a mixture of different noise types \cite{Chang2020Toward}.

\subsubsection{Response}
From the viewpoint of the detector response, we classify the stripe noise into additive and multiplicative stripes [Fig. \ref{Category}(a)]. The main difference between the additive stripe (${\bf{H}} = {\bf{X}} + {\bf{S}}$, where {\bf{H}} is the striped image, {\bf{X}} is the clean HSI, and {\bf{S}} is the stripe) and the multiplicative stripe (where ${\bf{H}}$ is modeled as the element-wise product between ${\bf{X}}$ and ${\bf{S}}$), is that the former is signal-independent whereas the latter is signal-dependent. We can observe that the intensities of the additive stripe along the stripe are usually close to a constant value. On the contrary, the intensity of the multiplicative stripe is highly associated with the image content. Normally, the additive model can be well applied to the multiplicative case after applying the logarithm.

\subsubsection{Intensity}
The stripe noise can be classified into normal-intensity stripes and deadlines [Fig. \ref{Category}(b)]. The deadlines are usually observed in HSIs due to the complete failure of the corresponding detectors. The intensity of the deadlines is zero which does not provide any useful information about the scene anymore. Compared with the normal-intensity stripe, the deadlines are more difficult to be recovered since no useful information can be utilized from the degraded pixels.

\subsubsection{Angle}
The stripe noise can be divided into vertical/horizontal and oblique ones [Fig. \ref{Category}(c)]. The stripe should be horizontal or vertical based on its imaging principle. However, in the subsequent remote sensing product, the geometric registration causes oblique stripes. Numerous destriping methods take advantage of the horizontal or vertical property of the stripe noise while such a valuable property no more holds for the oblique stripes. This makes the oblique stripes much more difficult to remove, given that the angle estimation must also be considered.

\subsubsection{Periodicity}
We classify stripe noises into periodic and non-periodic ones [Fig. \ref{Category}(d)]. The periodicity of the stripe mainly depends on the scanning mechanism of imaging instruments. The periodic stripes always appear in whisk-broom imaging systems where the sensors are placed along-track and scan forward and reverse across-track. A few examples of such imaging systems are the Moderate Resolution Imaging Spectroradiometer (MODIS), Landsat Thematic Mapper (TM), and MultiSpectral Scanner (MSS), while the non-periodic stripes exist in push-broom imaging systems due to its straight-line scan type such as the Satellite Pour l'Observation de la Terre (SPOT). As for the periodic stripe, it periodically locates in certain lines, which makes it easier to identify in the frequency domain. On the contrary, the non-periodic stripe is much more difficult to remove.

\subsubsection{Proportion}
The stripe noise can be divided into partial and full cover [Fig. \ref{Category}(e)]. The clean content in the case of a partial cover is a very beneficial clue to estimate the intensity of the striped regions using, for example, interpolation-based destriping methods. On the contrary, in fully covered stripe images, there is no such clue, which makes this case very difficult to handle.

\subsubsection{Length}
Stripe noise can be classified into full length and broken ones [Fig. \ref{Category}(f)]. The broken stripes mean that the length of the stripe lines is shorter than the image row, while the full-length stripes run through the whole image. Compared with the full-length stripes, it is difficult to precisely detect the location of the broken stripes. The line-pattern image structures are very similar to broken stripes from a local perspective.

\subsubsection{Width}
We can divide the stripe noise into single and broad ones  [Fig. \ref{Category}(g)]. The single-width stripe can be easily identified and well removed by various destriping methods. The broad stripe means that several adjacent detectors fail simultaneously within a relatively large region, which downgrades the performance of the gradient-based destriping methods dramatically. The full-covered stripe can be regarded as a special and difficult broad stripe.

\subsubsection{Mixed Noise}
The stripe noise always co-exists with the random noise \cite{Chang2020Toward}. 
The statistical distribution of the mixed noise is \textcolor{black}{more} complex due to the non-independent, non-identical property, and it is difficult to formulate the mixed noise with an explicit expression. The conventional destriping methods usually resort to the spectral correlation to remove the mixed noise. 

\begin{figure*}[htbp]
\begin{center}
    \includegraphics[width=0.99\textwidth]{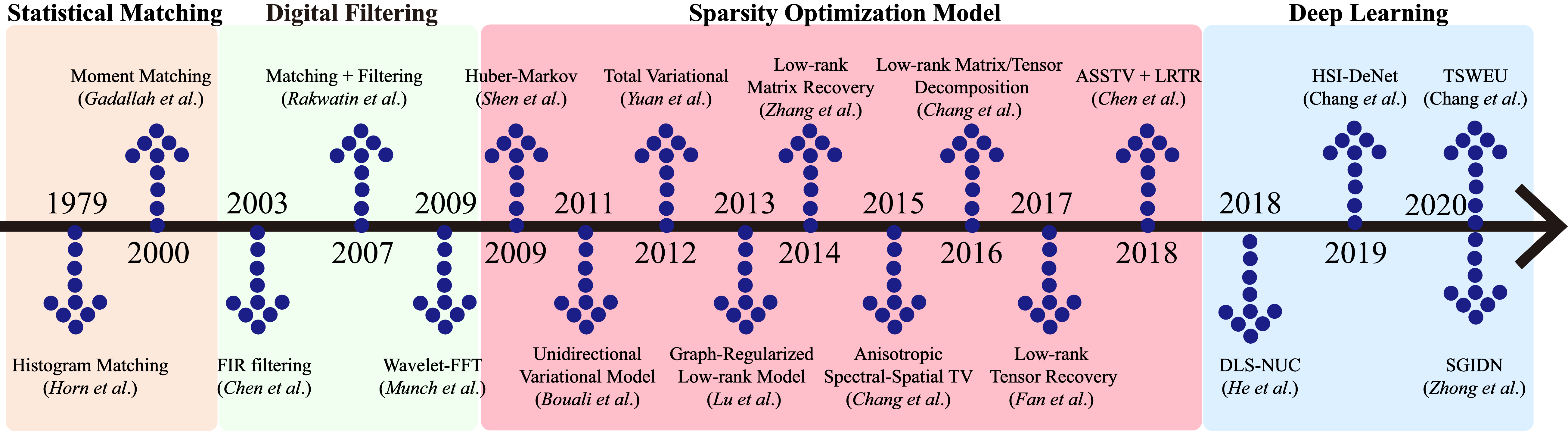}
 \end{center}
   \caption{Milestones of remote sensing image destriping methods: statistical matching, digital image processing/filtering, sparsity optimization models, and deep learning-based CNN methods. Before 2000, the typical methods were matching-based approaches. In the first decade of the 21st century, digital image processing methods occupied the mainstream, especially wavelet filtering. The time period from 2009 up to 2018 was dominated by sparsity optimization methods based on handcrafted features, including the 1D-gradient-based variational model, 2D low-rank matrix recovery, and the 3D low-rank tensor approximation. After 2018, HSIs destriping methods enter into a period of data-driven deep learning approaches.}
\label{Destriping Milestones}
\end{figure*}

\subsection{Conventional Techniques}
We categorize the conventional HSIs destriping methods into three general approaches: statistical matching, digital image processing/filtering, and a sparsity optimization model. The milestones of the HSIs destriping methods are presented in Fig. \ref{Destriping Milestones}.

\subsubsection{Statistical Matching}
The statistical matching methods were popular before 2000. The key idea of the statistical matching-based methods is to rectify the distribution of the degraded region to that of the clean reference. These methods, in fact, implicitly take advantage of the structural property of the stripes: vertically/horizontally located on certain regions while other regions are intact. Normally, the matching-based methods include two main steps: looking for the reference lines and hand-crafted statistical feature matching.

Horn \emph{et al.} \cite{horn1979destriping} utilized the histogram as the statistical feature for Landsat image stripe removal. Wegener \cite{wegener1990destriping} followed a similar idea and improved the searching strategy for the reference lines by homogeneous regions. Gadallah \emph{et al.} \cite{gadallah2000destriping} replaced the histogram matching with the moment feature, which improved the robustness with respect to the similarity assumption. Although the statistical matching methods can alleviate the striping artifact, they are less robust to the challenging categories of stripes, such as the fully-covered stripes and broken stripes, where both clean and striped regions are difficult to recover.

\subsubsection{Digital Image Processing/Filtering}
Numerous digital image processing-based destriping methods were proposed in the first decade of the 21st century. The key idea of the filtering-based methods is to identify the stripe component in the transformed frequency domain, not in the image domain. That is to say, the filtering-based methods treat the stripes as a special kind of noise, and formulate the destriping task as a denoising problem.

The finite-impulse response filter (FIR) has been extensively utilized for HSIs destriping \cite{simpson1998improved}, which mainly includes the following three steps: FFT transformation, striping component detection in frequency, and filter design for stripe component truncation. In 2009,  M\"{u}nch \emph{et al}. \cite{munch2009stripe} took the direction of the stripe into consideration via the wavelet, and the wavelet-based FFT method leading to impressive destriping performance. Hybrid approaches combining statistical matching with the filtering methods have also been presented \cite{rakwatin2007stripe, jung2010detection, jia2019destriping, Cui2020Multiscale}. It is worth noting that the filtering-based methods are suitable for  regular stripes, such as periodical stripes where the frequency can be easily separated from the image structures.

\subsubsection{Sparsity Optimization Approaches}
The sparsity optimization methods have undoubtedly dominated the HSIs destriping field in last decade. The core idea of the optimization methods is to formulate the destriping task as an ill-posed inverse problem. These methods can be derived from the \emph{maximum-a-posterior} (MAP) or the regularization framework in which the loss functional typically \textcolor{black}{includes a data-fidelity term and prior terms for both the image and stripe as follow:
\begin{equation}\label{eq:Destriping Cost}
\{\hat {{\bf X}}, \hat {{\bf S}}\}=\arg\min_{{\bf X}, {\bf S}}~\frac{1}{2}\left\| {\bf H}- {\bf X} - {\bf S}\right\|^{2}_{F} + \lambda\phi({ \bf X}) + \rho\psi({ \bf S}).
\end{equation}
Note that this model is not only applicable for additive stripe, but also other stripes. For example, the multiplicative stripe can be simply transformed to the additive stripe with the logarithm function.} According to the categories of the priors, optimization-based methods can further be divided into 1D gradient-based variational models\cite{shen2009map, chang2013robust, chang2014simultaneous, chang2015anisotropic, liu2016stripe, liu2018universal, huang2020joint}, 2D low-rank matrix recovery methods \cite{acito2011subspace, lu2013graph, chang2016remote, chang2017transformed, chen2017group}, and 3D low-rank tensor approximations \cite{fan2017hyperspectral, chen2018destriping, cao2018destriping, liu2019wavelet, chang2020hyperspectral}.


Many destriping methods proposed in the literature exploited spectral gradient to capture the spectral correlation \cite{HyMiNoR}. In 2009, Shen \emph{et al}. \cite{shen2009map} firstly employed the MAP framework and introduced the Huber-Markov prior for destriping which can adaptively preserve the edge and detail information. To better model the directionality of the stripe, Bouali \emph{et al}. \cite{bouali2011toward} proposed the delicate unidirectional variational model where the gradients across the stripes are minimized while the gradients along the stripes are well preserved. The latter work further takes the spectral information into consideration, instead of the single image destriping. Naturally, Chang \emph{et al}. \cite{chang2015anisotropic} extended the unidirectional variational model \cite{bouali2011toward} to the anisotropic spectral-spatial TV (ASSTV) model for multispectral image destriping. It is worth noting that the aforementioned methods aim at estimating the clean image directly. An alternative way is to estimate the stripe component instead \cite{carfantan2010statistical} since the structure of the stripe is much simpler than that of the image.

The 2D low-rank matrix recovery methods could better preserve the spatial or spectral correlations, which have been naturally introduced for HSIs destriping \cite{acito2011subspace, lu2013graph, chang2016remote, chang2017transformed, chen2017group}. The low-rank-based destriping methods mainly utilize the spectral correlation and perform a robust principal component analysis (RPCA) \cite{wright2009robust} so as to decouple the sparse stripe error from the image component. For example, Lu \emph{et al}. \cite{lu2013graph} lexicographically ordered the 3D HSI into a 2D matrix, and applied the RPCA on the constructed low-rank matrix to remove the stripe noise. Instead of enforcing the low-rank constraints on the image, Chang \emph{et al}. \cite{chang2016remote} argued that the low-rank property of the stripe component is much better than that of the image, and the low-rank image decomposition method has achieved state-of-the-art destriping performance.

Although the vector/matrix-based methods have achieved excellent destriping results, they may cause spectral and spatial distortion. To alleviate this issue, low-rank tensor recovery methods have emerged in recent years \cite{fan2017hyperspectral, chen2018destriping, cao2018destriping, liu2019wavelet, chang2020hyperspectral}. For example, Fan \emph{et al}. \cite{fan2017hyperspectral} proposed the low-rank tensor recovery model for mixed Gaussian and sparse stripe noise removal. Chen \emph{et al}. \cite{chen2018destriping} presented a tensor image decomposition framework where the $\ell_{2,1}$ norm is used to accommodate the column-wise group sparsity of the stripe. Overall, these sparsity optimization-based methods are explainable and effective for all kinds of stripes. However, compared to supervised deep learning-based algorithms such as (CNN) and the filtering-based methods, they are computationally expensive and, therefore, not suitable for real-time applications. 

\begin{table*}[htbp]
\footnotesize
\centering
\renewcommand{\arraystretch}{1.05}
\caption{Quantitative results of different methods under different stripe categories and levels. }
\label{Destriping Quantitative}
\begin{tabular}{|c|c|c|c|c|c|c|c|c|c|}
\hline
Category                                                                                         & Index & Stripe & WFAF \cite{munch2009stripe}  & TV  \cite{rudin1992nonlinear}   & UTV  \cite{bouali2011toward}  & SLD \cite{carfantan2010statistical}   & LRSID \cite{chang2016remote} & DLS-NUC \cite{he2018single}& TSWEU \cite{Chang2020Toward} \\ \hline
\multirow{4}{*}{\begin{tabular}[c]{@{}c@{}}Multiplicative\\ {[}0.7 1.3{]}\end{tabular}}          & PSNR  & 8.64   & 35.17  & 29.12  & {36.44}  & {43.87 } & 36.06  & 33.33   & \bf{46.96}  \\ \cline{2-10}
                                                                                                 & SSIM  & 0.5124 & 0.9768 & 0.8526 & 0.9851 & {0.9959} & {0.9852} & 0.9626  & \bf{0.9977} \\ \cline{2-10}
                                                                                                 & MICV  & 22.52  & 10.35  & \bf{92.40}  & 9.94   & {12.70}  & 11.67  & 12.45   & {12.53}  \\ \cline{2-10}
                                                                                                 & MMRD  & 0      & 0.057  & 0.094  & 0.056  & {0.048}  & {0.052}  & 0.072   & \bf{0.040} \\ \hline
\multirow{4}{*}{\begin{tabular}[c]{@{}c@{}}Periodic\\ {[}-20 20{]}\end{tabular}}                 & PSNR  & 23.75  & {33.39}  & 26.99  & 31.63  & {34.73}  & 32.66  & 32.80   & \bf{39.78}  \\ \cline{2-10}
                                                                                                 & SSIM  & 0.5569 & {0.9607} & 0.7452 & 0.9465 & {0.9703} & 0.9575 & 0.9540  & \bf{0.9910} \\ \cline{2-10}
                                                                                                 & MICV  & 3.17   & 11.67  & 6.38   & 11.14  & {13.08}  & \bf{19.15}  & 11.71   & {12.18}  \\ \cline{2-10}
                                                                                                 & MMRD  & 0      & 2.25   & \bf{1.49} & 2.32   & 2.22   & {2.09}   & 2.18    & {2.12}   \\ \hline
\multirow{4}{*}{\begin{tabular}[c]{@{}c@{}}Non-periodic\\ {[}-30 30{]}\end{tabular}}             & PSNR  & 23.72  & 33.67  & 26.98  & 31.60  & {39.73}  &{ 34.31}  & 33.07   & \bf{41.86}  \\ \cline{2-10}
                                                                                                 & SSIM  & 0.5655 & 0.9650 & 0.7521 & 0.9474 & {0.9926} & {0.9740} & 0.9550  & \bf{0.9942} \\ \cline{2-10}
                                                                                                 & MICV  & 3.20   & 9.77   & 6.68   & 8.39   & {12.71}  & \bf{14.51}  & 12.61   & {13.51}  \\ \cline{2-10}
                                                                                                 & MMRD  & 0      & {0.8075} & \bf{0.5851} & {0.8102} & 0.8636 & 0.8379 & 0.8802  & 0.8975 \\ \hline
\multirow{4}{*}{\begin{tabular}[c]{@{}c@{}}Broken\\ {[}-40 40{]}\end{tabular}}                   & PSNR  & 32.59  & 34.26  & 27.09  & {37.02}  & {34.50}  & 33.27  & 33.02   & \bf{49.05}  \\ \cline{2-10}
                                                                                                 & SSIM  & 0.9163 & 0.9414 & 0.7524 & {0.9771} & {0.9520} & 0.9471 & 0.9388  & \bf{0.9983} \\ \cline{2-10}
                                                                                                 & MICV  & 7.96   & 9.93   & {11.13} & 10.06  & 11.04  & \bf{18.45}  & 9.14    & {12.75}  \\ \cline{2-10}
                                                                                                 & MMRD  & 0      & \bf{0.1202} & 0.2890 & 0.1479 & {0.1242} & 0.1808 & 0.1964  & {0.1213} \\ \hline
\multirow{4}{*}{\begin{tabular}[c]{@{}c@{}}Mixed Noise\\ {[}-20 20{]}\\ Sigma = 10\end{tabular}} & PSNR  & 24.79  & 27.49  & {28.42}  & 27.50  & 27.83  & 28.01  & {28.90}   & \bf{33.24}  \\ \cline{2-10}
                                                                                                 & SSIM  & 0.5879 & 0.6989 & {0.7855} & 0.6971 & 0.7035 & 0.7284 & {0.7638}  & \bf{0.9256} \\ \cline{2-10}
                                                                                                 & MICV  & 3.42   & 4.26   & {6.12}   & 4.23   & 4.26   & 4.49   & {5.25}    & \bf{13.89}  \\ \cline{2-10}
                                                                                                 & MMRD  & 0      & {0.6118} & 0.6155 & \bf{0.5347} & 0.6489 & {0.6051} & 0.7225  & 0.8442 \\ \hline
\end{tabular}
\end{table*}

\subsection{Deep Learning-based Techniques}

Deep learning-based methods such as CNNs have been extensively used for various computer vision tasks. The advantage of CNNs over the conventional methods for destriping is two-fold. First, thanks to the universal approximation theory \cite{hornik1989multilayer}, which theoretically proves that CNN could implicitly approximate any complicated distribution of arbitrary mixed noise, the supervised CNN could well handle arbitrary stripe noise. Second, due to the simple operation of the network, its test phase is extremely fast which makes it quite suitable for real-time application.

In 2017, a three-layer CNN was firstly proposed by Kuang \emph{et al}. \cite{kuang2017single} to remove the stripe noise in infrared images, namely the non-uniform correction. Xiao \emph{et al}. \cite{xiao2018removing} increased the depth of the network so as to improve the network representation ability. Further, residual learning has been employed by Xiao et al. \cite{xiao2018removing} for better destriping. Moreover, Chang \emph{et al}. \cite{Chang2020Toward} introduced a multi-scale strategy into the network to enhance the representation. Recently, Chang \emph{et al}. \cite{Chang2020Toward} embedded the wavelet into a two-stream CNN framework to learn the internal directional property of the stripe. A unique CNN architecture design by Zhong \emph{et al}. \cite{Zhong2020Satellite} was introduced, which integrates 2D and 3D convolutions, residual learning, and supplementary gradient channels to capture intrinsic spectral-spatial features in HSIs and the unidirectional property of stripe. Most recently, to overcome the lack of paired training data for deep learning, Song \emph{et al}. \cite{Song2020Unsupervised} proposed a novel unsupervised HSIs destriping method with subband cycle-consistent adversarial network.

\begin{figure*}[htbp]
\begin{center}
    \includegraphics[width=0.99\textwidth]{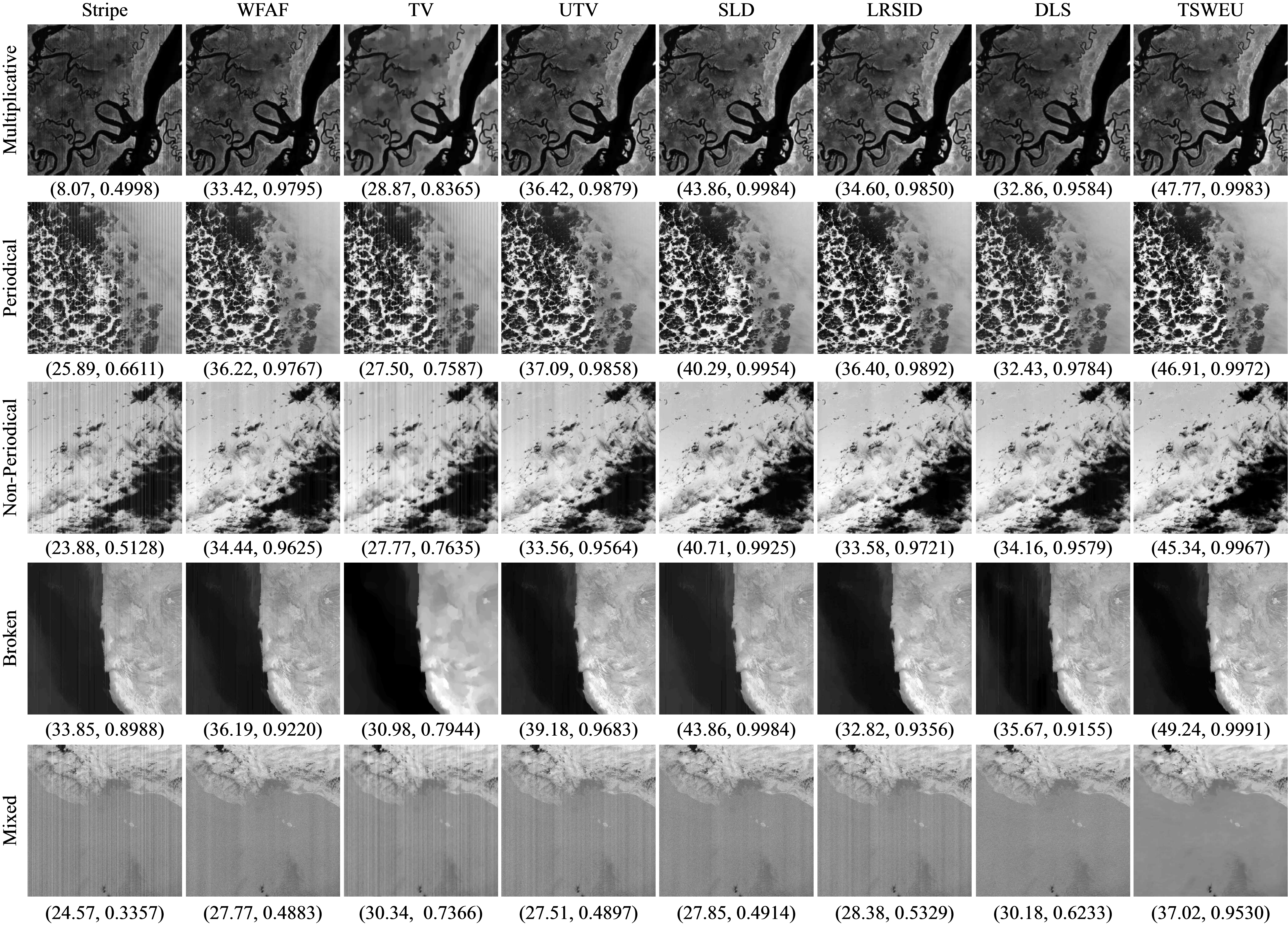}
 \end{center}
   \caption{Visual destriping results of competing state-of-the-art methods for different stripe categories and stripe levels. The corresponding (PSNR, SSIM) values are listed below the images.}
\label{Visual Destriping}
\end{figure*}

\subsection{Results and Comparisons}

We evaluate the \textcolor{black}{destriping methods with the state-of-the-art performance in each category} including \textcolor{black}{the filtering-based method}: wavelet Fourier adaptive filter (WFAF) \cite{munch2009stripe}, \textcolor{black}{the optimization-based methods}: TV \cite{rudin1992nonlinear}, unidirectional TV (UTV) \cite{bouali2011toward}, statistical linear destriping (SLD) \cite{carfantan2010statistical}, low-rank single-image decomposition (LRSID) \cite{chang2016remote}, \textcolor{black}{the deep learning-based methods}: deep learning stripe (DLS) \cite{he2018single},  two-stream wavelet enhanced U-net (TSWEU) \cite{Chang2020Toward}. As for the dataset, we use the 15 remote sensing images collected in \cite{Chang2020Toward} for the testing. The full-reference indices: PSNR and SSIM, and the no-reference assessments: inverse coefficient of variation (ICV) \cite{bouali2011toward} and mean relative deviation (MRD) \cite{shen2009map} are employed for the quantitative assessment. 

\textcolor{black}{For the simulation of the stripes, different stripes have different simulation procedures. We take the classical additive stripe as an example. We generate the striped image {\bf{H}} by adding the stripes {\bf{S}} to the clean image {\bf{X}}. The matrix coefficient is generated by $repmat((stripe_{max} - stripe_{min}).*rand(1,size(H,2)) + stripe_{min},size(H,1),1)$. $stripe_{max}$ and $stripe_{min}$ are the pre-defined hyper-parameters which determine the range of the additive stripe. “repmat” and “size” are the Matlab functions. Interested readers could refer to the released simulating codes for other stripe categories.}

\subsubsection{Quantitative Evaluation}
We show that the quantitative results of different competing methods over different stripe categories and stripe levels. We choose the most representative and typical stripes: multiplicative, periodical, non-periodical, broken, and mixed. It is worth noting that the additive, covered, and full are also included in the listed five cases. The deadlines and oblique stripes are the only two kinds of stripes that are not tested here since very few methods can handle such degradation. From Table \ref{Destriping Quantitative}, we have the following observations. First, the deep learning-based methods can consistently obtain the best performance among all the methods. Second, SLD could well handle the regular stripes since its rank-1 constraint provides a satisfactory fit for these regular stripes. However, for the broken or mixed noise where the assumption does not hold, the performance of SLD would drop very rapidly. Third, the filtering-based WFAF achieves impressive performance for the periodic stripes, since the periodic stripes have very regular high-frequency components in the frequency domain.

\textcolor{black}{Moreover, in Table \ref{Destriping Running Time}, we report the running times under different image sizes. We perform the experiments on MATLAB 2017a, with an Intel i7 CPU at 3.6 GHz, an NVIDIA GEFORCE RTX 1080Ti GPU, and 32-GB memory. The training and test times are considerably different from each other. We can observe that the running time of the learning based methods including DLS and TSWEU are significantly faster than that of other methods. Note that, the learning-based methods need to be trained on large-scale datasets, which means that they need hours to days for training, while the inference/testing time is usually very fast (within a second). On the contrary, the hand-crafted methods do not require the training procedure, while the testing time is relatively slow (typically one or two orders of magnitude slower than deep neural networks).} 


\subsubsection{Qualitative Evaluation}
We show several visual destriping results of the competing methods in Fig. \ref{Visual Destriping}. The main observations are: (1) Most of the existing methods work well for typical stripes such as the periodical/non-periodical and multiplicative/additive stripes. However, for the broken stripes where the low-rank property no longer holds for the stripes, most of the existing methods fail to satisfactorily restore the image. (2) For the mixed stripes and random noise, most of the existing single image-based destriping methods can not handle random noise well. Additionally, random noise may even bring a negative impact to stripe removal. (3) The robustness and visual results of the learning-based TSWEU have consistently obtained the best performance among the competing methods. It is worth noting that the TSWEU can handle more kinds of stripe such as deadlines and oblique stripes.

\subsubsection{\textcolor{black}{Discussion}}
\textcolor{black}{In general, the hand-crafted-based methods are designed for certain kinds of the stripe, such as the smoothness \cite{rudin1992nonlinear}, low-rank \cite{chang2016remote}, and directionality \cite{bouali2011toward}. That is to say, these methods are usually limited to certain stripes. The learning-based methods heavily depend on the training datasets. Once we have comprehensively defined the stripe categories, the deep learning destriping methods could consistently obtain better performance.
} 

\begin{table}[t]
\centering
\renewcommand{\arraystretch}{1.3}
\textcolor{black}{\caption{Processing time of the competing destriping methods under different image sizes.}
\label{Destriping Running Time}
\begin{tabular}{cccccccc}
\toprule
Size      & WFAF  & TV     & UTV    & SLD   & LRSID   & DLS & TSWEU \\ 
\midrule
$128^2$   & 0.02 & 0.15  & 0.14  & 0.03 & 1.07   & 0.02   & 0.05 \\ 
$256^2$  & 0.04 & 0.65  & 0.56  & 0.07 & 2.76   & 0.04   & 0.06 \\ 
$512^2$   & 0.11 & 3.31  & 2.99  & 0.19 & 10.81  & 0.05   & 0.07 \\ 
$1024^2$ & 0.41 & 14.19 & 13.55 & 0.64 & 49.57  & 0.14   & 0.11 \\ 
\bottomrule
\end{tabular}}
\end{table}

\subsection{Remaining Challenges}
\textcolor{black}{Although the existing destriping methods have made significant progress in recent years, there remain many challenges that need to be solved in the future.}

\subsubsection{Huge Gap between Simulated and Real Degradation}
\textcolor{black}{Based on the degradation model, the optimization methods solve the corresponding ill-posed problem, while deep learning methods simulate a training set of clean and corrupted images. That is to say, the destriping performance of the optimization and deep learning-based methods heavily rely on the accuracy of the degradation model. However, in real cases, the physical degradation procedure of the stripe is very complex. The existing simplified additive model or multiplicative model cannot well accommodate all kinds of stripes. The huge gap between simulated and real degradation becomes a serious impediment to the development of the destriping field. In our opinion, it is difficult to accurately figure out the mathematical degradation model. To address this problem, we may rely on the unsupervised/self-supervised CNN frameworks (please refer to the definitions in Section III.B.) from a data-driven perspective. In the future, more works will be expected in this direction to reduce the gap.}

\subsubsection{Generalization to Different Stripes}
\textcolor{black}{As we have mentioned in the previous subsections, there are at least 16 kinds of stripes in HSIs. Most of the previous methods are limited to certain stripes. For example, the multiplicative and the oblique stripes are rarely considered. Moreover, the stripe artifacts are ubiquitous in various imaging systems, such as the non-uniformity in infrared focal plane array \cite{chang2019Infrared} or scanning electron microscope images \cite{fehrenbach2012variational}. Therefore, how to accommodate the trained model for general stripe noises including different directions, widths, lengths, and so on, is a crucial issue for practical applications. A natural idea is to train a powerful network with very large-scale and diverse training samples. We argue that it may be more reasonable to make efforts by embedding the physical prior knowledge of the stripe (such as line-pattern) into the network, so as to improve the generalization of different stripes.}

\subsubsection{Evaluation of the Destriping Performance}
\textcolor{black}{Evaluation indexes are used to measure whether the proposed destriping method is effective or not. The existing assessments are mainly based on well-known spatial PSNR/SSIM and spectral ERGAS/SAM. These reference-based destriping assessments are used to measure the similarity between the destriping results and the original clean images, and also how many distortions have been unexpectedly introduced. 
}


\section{Hyperspectral Deblurring}
HSIs are often blurred during the acquisition process due to a fundamental limitation of the system or the atmospheric turbulence. These blurring artifacts unexpectedly suppress the high-frequency component of the textures/edges and, thus, weaken the discriminative features of HSIs. Therefore, the goal of HSIs deblurring is to recover a sharp image from a blurred one, which is a crucial step in improving the resolution for subsequent applications. Mathematically, the HSIs blur degradation process is often formulated as follows:
\begin{equation}\label{eq:Blur_Degradation}
{\bf H} = {\bf B}{\bf{X}} + {\bf{N}},
\end{equation}
where ${\bf H} \in \mathbb{R}^{n\times p}$ is an observed HSI, ${\bf X} \in \mathbb{R}^{n\times p}$ represents the desired clean HSI, ${\bf {{B}}} \in {\mathbb{R}^{n \times n}}$ denotes the blur operator also known as point spread function (PSF), ${\bf {{N}}} \in {\mathbb{R}^{n \times p}}$ is the Gaussian noise. \textcolor{black}{Here, we assume the blur operator is constant throughout the spectral bands. In general, the blur operator might vary throughout the spectral bands. Thus the deblurring procedure can be mathematically formulated as:
\begin{equation}\label{eq:Deblurring Cost}
\{\hat {{\bf X}}, \hat {{\bf B}}\}=\arg\min_{{\bf X}, {\bf B}}~\frac{1}{2}\left\| {\bf H}-{\bf B}{\bf X}\right\|^{2}_{F} + \lambda\phi({ \bf X}) + \rho\psi({ \bf B}),
\end{equation}
where $\lambda$ and $\rho$ are the regularization parameters. The goal is to estimate the clean image {\bf X} and blur kernel {\bf B} from the given blur images {\bf H} with proper prior terms $\phi({ \bf X})$ for the clean image and $\psi({ \bf B})$ for blur kernel respectively. Next, we will briefly analyze the typical blur kernel ${\bf{{B}}}$ and the representative priors for ${\bf{{X}}}$.}


\begin{figure}[t]
\begin{center}
    \includegraphics[width=0.5\textwidth]{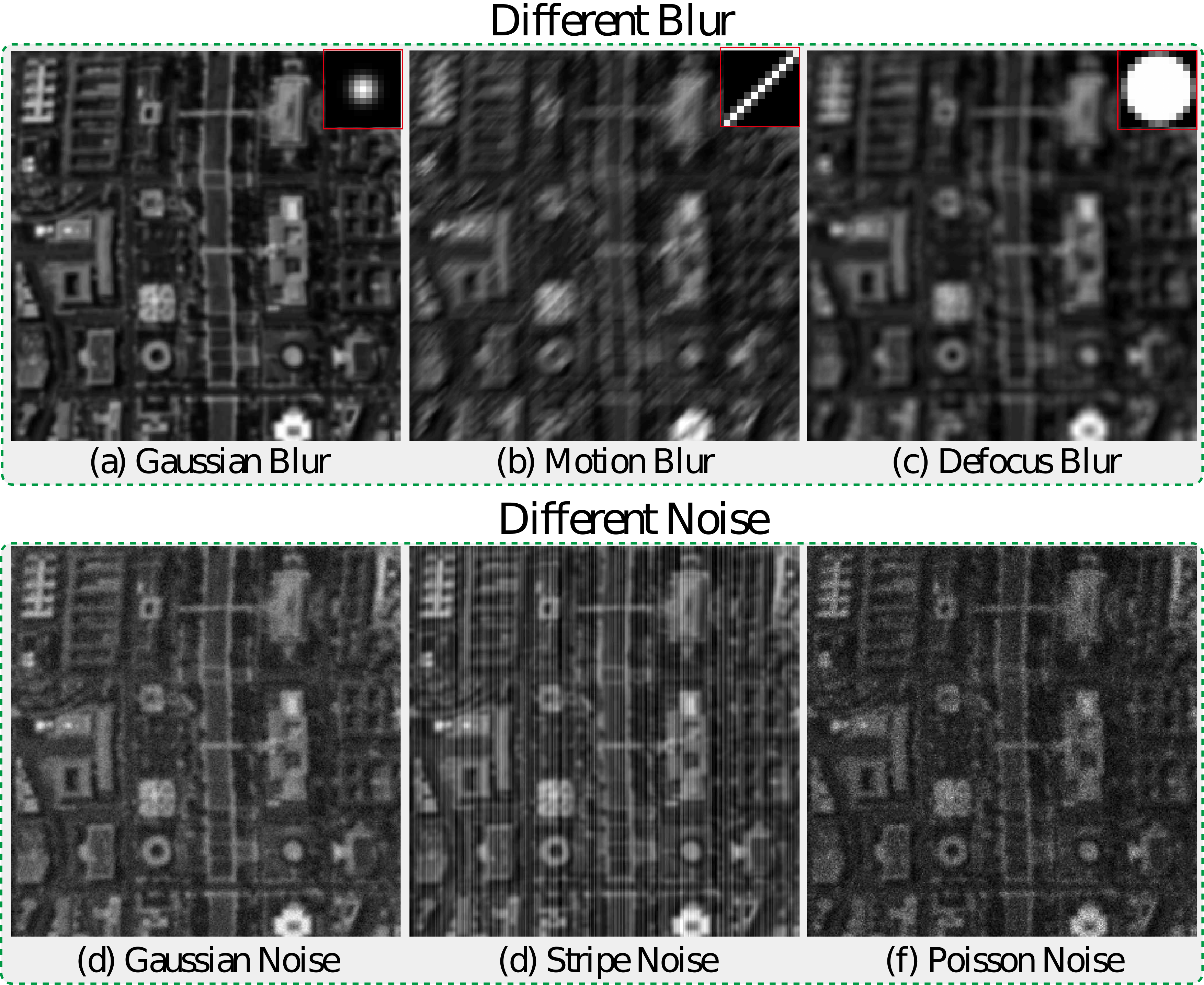}
 \end{center}
  \caption{Illustration of different blurs and the co-existing noise. The top row shows three different blurs: Gaussian blur, motion blur, and defocus blur. The bottom row shows three representative noises in presence of blur: Gaussian noise, stripe noise, and Poisson noise.}
\label{Blur Category}
\end{figure}

\begin{figure*}[htbp]
\begin{center}
    \includegraphics[width=0.99\textwidth]{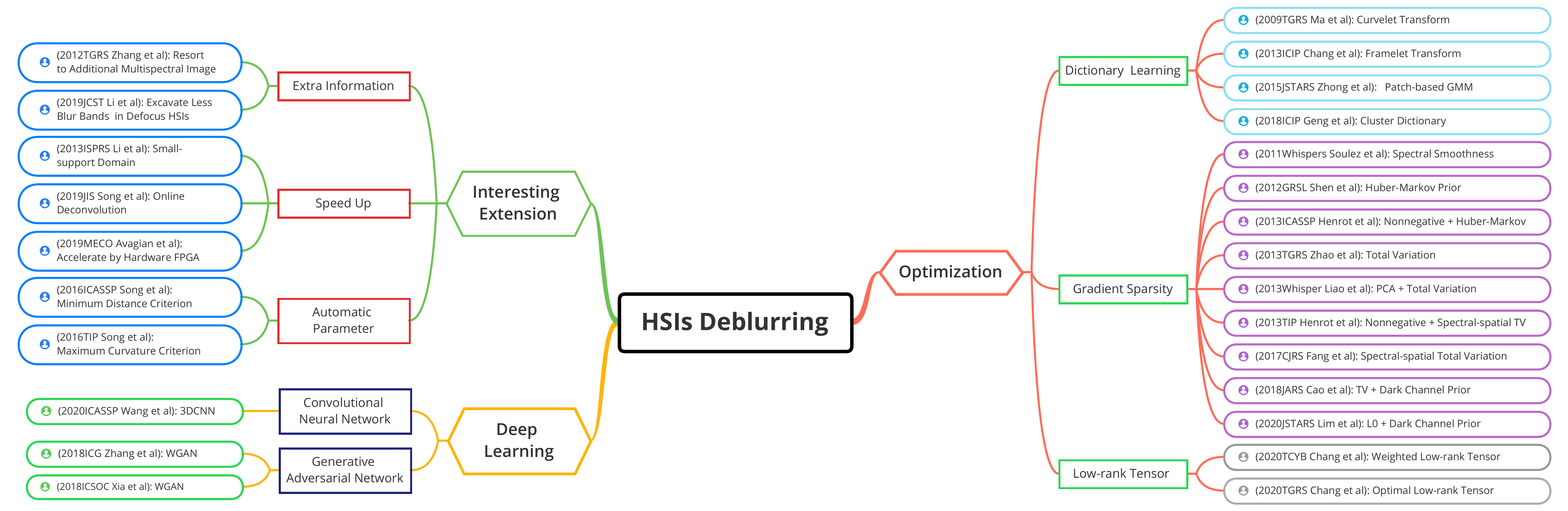}
 \end{center}
   \caption{Taxonomy of the main HSIs deblurring methods.}
\label{Deblurring Evolution}
\end{figure*}

\subsection{HSIs Blur Kernel and Noise}
There are many blur factors in the imaging procedure: atmospheric turbulence, vibrations of the imaging system platform, the diffraction limit, and the lens being out-of-focus. Consequently, the blur in HSIs can be classified into three main categories: turbulence blur, motion blur, and defocus blur. We show the simulated blurring HSI and the corresponding blur kernel in Fig. \ref{Blur Category}.

\subsubsection{Turbulence Blur}
For the Earth observation satellite-borne HSI, the light would pass through several kilometers of the atmosphere. Unfortunately, the atmosphere can not be simply regarded as a uniform transmission medium since the density, humidity, and temperature change the refractive index along the long-distance optical transmission path. Consequently, atmospheric turbulence brings geometric distortions  and time-varying blur to the final images, namely the turbulence blur. In this work, we simplify the atmospheric turbulence in HSIs as a Gaussian blur as shown in Fig. \ref{Blur Category}(a). It is worth noting that the real atmospheric turbulence is much more complex than a Gaussian blur. 

\subsubsection{Motion Blur}
The motion blur is caused by the relative motion between the sensor and the scene/object in a single exposure. Satellite platforms are usually moving rapidly and the push-broom/whisk-broom imaging systems are based on the scanning mechanism by sacrificing the temporal resolution so as to obtain 3D spatial-spectral information. Thus, system vibrations or motion would unexpectedly bring motion blur and the corresponding blur reflects the motion trajectory of the sensor [Fig. \ref{Blur Category}(b)].

\subsubsection{Defocus Blur}
The defocus blur refers to the translation of the focus of the scene along the optical axis away from the imaging sensor. The main reason is that different wavelengths in the different bands do not focus on the same focal plane \cite{li2019defocus}. Defocus blur can be modeled by a uniform disk, and the defocused image significantly loses sharp edges and details [Fig. \ref{Blur Category}(c)].

In HSIs, blur and noise are common degradation issues. We have analyzed the characteristic of various noises in HSI denoising and destriping section. Here, we show the typical noise such as the Gaussian noise, stripe noise, and Poisson noise, coexisting with the blurs as shown in Fig. \ref{Blur Category}(d)-(f). The noise in the deblurring procedure would significantly increase the ill-posedness and difficulty to design stable estimators of the solution. Numerous deblurring methods have been proposed to handle the HSIs deblurring problem. In the next subsection, we review the existing HSIs deblurring methods from two main categories: optimization and learning paradigms.

\subsection{HSIs Deblurring Paradigm}
Natural image deblurring aims to recover a sharp latent image from a blurred one \cite{Qi2008High}, which is a classical and active research field in the last decades. In this survey, we introduce the existing HSI deblurring methods from optimization and deep learning perspectives, \textcolor{black}{as shown in Fig. \ref{Deblurring Evolution}}.

\subsubsection{Sparsity Optimization Models}
The key idea of the sparsity optimization-based methods is to formulate the HSI deblurring as an ill-posed inverse problem. Normally, an energy function is constructed based on the \emph{maximum-a-posterior} or regularization framework which reflects the modeling characteristic of the noise, the HSIs prior, and the blur kernel. Then, we need to apply the optimization algorithms to iteratively minimize the constructed functional.

Any small perturbation of the observed image, such as the noise may cause large distortions in the solution. Therefore, the prior/regularization is necessarily enforced so as to guarantee the solution's stability. Most of the HSI deblurring methods focus on designing sophisticated prior terms for HSIs. Here, we further classify the existing optimization-based HSI deblurring methods into three categories: dictionary learning, gradient sparsity, and low-rank tensor methods.

Dictionary learning such as K-SVD \cite{Aharon2006K} has been widely used in computer vision for image restoration \cite{Elad2006Image}. The dictionary learning aims at finding a sparse representation of the HSIs $\textbf{\emph{X}} $ via a linear combination of basic elements:
\begin{equation}\label{eq:Dictionary learning}
\setlength{\abovedisplayskip}{2pt}
\setlength{\belowdisplayskip}{2pt}
{\bf{{X}}} = {\bf{{D}}}{\boldsymbol{\alpha}} ,
\end{equation}
where $\bf{{D}}$ is called the dictionary, ${\boldsymbol{\alpha}}$ is the sparse code (i.e., dictionary coefficients). One key issue in dictionary learning is to determine the basis of the dictionary $\bf{{D}}$. Ma \emph{et al}. \cite{Ma2009Deblurring} introduced the compressed sensing theory with highly incomplete measurements, and applied the curvelet threshold for the HSIs deblurring. Chang \emph{et al}. \cite{chang2013joint} employed the framelet transform for remote sensing image deblurring in presence of the stripe noise. 
Geng \emph{et al}. \cite{geng2018structral} proposed a structural compact core tensor dictionary learning model for multispectral image deblurring.

The gradient sparsity-based methods utilize smoothness in the spatial and spectral gradient domain. The most famous gradient-based method is TV \cite{rudin1992nonlinear}, which can well preserve sharp edges.  Soulez \emph{et al}. \cite{soulez2011restoration} proposed to utilize spectral smoothness along with spatial sparsity for HSIs restoration. Later, the spatial-spectral joint TV \cite{fang2017hyperspectral} was proposed to accommodate the 3D structure of the HSIs. Henrot \emph{et al}. \cite{henrot2012fast} incorporated the non-negative constraint into the spatial-spectral joint TV model for HSIs deblurring. Further, Henrot \emph{et al}. \cite{henrot2013edge} introduced the Huber-Markov variational model with spatially local adaptive edge-preserving ability for HSIs deblurring. From the spectral viewpoint, Cao \emph{et al}. \cite{cao2018dark} and Lim \emph{et al}. \cite{lim2020texture} presented the dark channel prior for HSIs deblurring, along with the $\ell_0$- and $\ell_1$-based TV regularizer 

In general, most previous HSI deblurring methods mainly exploit spatial or spectral information while \textcolor{black}{few}    of them have utilized the nonlocal self-similarity property in HSIs. Chang \emph{et al}. \cite{chang2020hyperspectral, chang2020weighted} proposed the low-rank tensor prior to model the spatial nonlocal self-similarity and spectral correlation property, simultaneously, to better preserve the intrinsic spectral-spatial structural correlation.

\begin{table*}[]
\caption{Quantitative results of different methods under different blur cases on CAVE dataset.}
\renewcommand{\arraystretch}{1.1}
\label{CAVE_Deblurring Quantitative}
\centering
\begin{tabular}{|c||c|c|c|c||c|c|c|c||c|c|c|c|}
\hline
\multirow{1}{*}{} & \multicolumn{4}{c||}{\begin{tabular}[c]{@{}c@{}}Gaussian Blur (8*8, Sigma = 3)\end{tabular}} & \multicolumn{4}{c||}{\begin{tabular}[c]{@{}c@{}}Motion Blur: Linear (Length = 45) \end{tabular}} & \multicolumn{4}{c|}{\begin{tabular}[c]{@{}c@{}}Uniform Blur: (s = 12)\end{tabular}} \\ \cline{1-13}
Method                        & PSNR                 & SSIM                  & ERGAS                & SAM                   & PSNR                  & SSIM                  & ERGAS                 & SAM                   & PSNR               & SSIM               & ERGAS               & SAM             \\ \hline
Blurred                        &  32.61                  & 0.9125                 & 135.80                  &0.0736                 &  25.35                   &  0.7888                &  297.98                  &  0.1403                & 29.62                 & 0.8588              & 187.50                 &   0.0924         \\ \hline
HL                              & 37.28                   & 0.9460                 & 83.88                    & 0.0676                &  31.78                   &   0.8430               &  146.30                  &  0.1801                & 35.11                 & 0.9163              & 104.82                 &    0.0887        \\ \hline
FPD                           & 38.84                   & 0.9617                 & 68.48                    & 0.0734                &  30.06                    &  0.8767               &   188.84                  &  0.1483               & 36.16                  & 0.9467              & 89.65                   &   0.0957       \\ \hline
SSTV                        & 37.61                    & 0.9527                 & 80.91                    & 0.0658                &  32.30                   &  0.8722                &   137.50                 &   0.1522               & 35.73                 & 0.9262              & 97.69                   &   0.0844        \\ \hline
WLRTR                    & 55.68                    & 0.9979                 & 9.96                      & 0.0250               &  61.77                    &   0.9992               &    4.75                    &  0.0213                & 53.74                 &  0.9965             & 11.63                   &   0.0541       \\ \hline
OLRT                       & \bf{57.02}                    & \bf{0.9984}                 & \bf{8.44}                      & \bf{0.0224}                & \bf{62.91}                    &   \bf{0.9992}               &    \bf{4.20}                    & \bf{0.0174}                 & \bf{57.84}                 & \bf{0.9985}              & \bf{7.64}                     &   \bf{0.0228}       \\ \hline
\multirow{1}{*}{} & \multicolumn{4}{c||}{\begin{tabular}[c]{@{}c@{}}Gaussian Blur (17*17, Sigma = 7)\end{tabular}} & \multicolumn{4}{c||}{\begin{tabular}[c]{@{}c@{}} Motion Blur: NonLinear \end{tabular}} & \multicolumn{4}{c|}{\begin{tabular}[c]{@{}c@{}}Uniform Blur: (s = 25)\end{tabular}} \\ \cline{1-13}
 Method                       & PSNR                 & SSIM                  & ERGAS                & SAM                   & PSNR                  & SSIM                  & ERGAS                 & SAM                   & PSNR               & SSIM               & ERGAS               & SAM                \\ \hline
Blurred                        &  28.69                 & 0.8428                 & 206.94                  & 0.1020                  &  22.14                   &  0.6984               &   423.94                 &    0.2391               &    26.18             &   0.7962            &   272.13               &    0.1293           \\ \hline
HL                              & 32.59                  & 0.8819                 & 137.14                  & 0.1075                  &  29.37                   &  0.7988               &   191.35                  &   0.2051               &    31.31             &   0.8460            &   157.30               &    0.1484           \\ \hline
FPD                            & 33.16                  & 0.9114                 & 125.11                  & 0.1163                  &  21.58                   &   0.6036              &   465.01                  &   0.2155               &   31.14              &   0.8816            &   161.31               &    0.1345           \\ \hline
SSTV                         & 33.08                  & 0.8944                 & 129.84                  & 0.0989                  &  28.56                  &   0.7928               &   207.29                  &   0.2147               &    31.68             &   0.8608            &   150.58               &    0.1345            \\ \hline
WLRTR                     & 49.42                  & 0.9926                 & 20.87                    & 0.0439                  & 54.46                   &   0.9971               &    11.37                   &   0.0307               &    49.91             &   0.9928            &   19.31                 &    0.0459          \\ \hline
OLRT                         & \bf{50.45}                  & \bf{0.9937}                & \bf{18.26 }                   & \bf{0.0387}                   & \bf{57.11}                   &   \bf{0.9982}               &    \bf{8.21}                    &   \bf{0.0241}               &    \bf{51.25}             &   \bf{0.9941}            &   \bf{16.48}                 &    \bf{0.0404 }         \\ \hline
\end{tabular}
\end{table*}

\subsubsection{Deep Learning Models}
Deep learning has been extensively used in natural image deblurring \cite{Nah2017Deep} and demonstrated its advantage over handcrafted regularizers. Consequently, Zhang \emph{et al}. \cite{zhang2018generative} proposed an end-to-end learnable method based on GANs for HSIs deblurring. However, the CNN-based models are only suitable for several specific types of blurs and have limits against more general spatially varying blurs. To address the generalization issue, benefiting from the variable splitting technique, the plug-and-play strategy was proposed to combine the optimization and the CNN. Wang \emph{et al}. \cite{wang2020learning} proposed a hyperspectral deconvolution technique that plugs a spectral-spatial 3D-CNN prior into the optimization framework, which could simultaneously obey the physical degradation and enjoy the powerful representation ability of CNN.

\subsection{Interesting Extensions}
\subsubsection{Joint with Other Tasks}
The blur always coexists with the noise, e.g., random noise or stripe noise \cite{liao2013hyperspectral}. 
In \cite{chang2013joint}, it was shown that the joint destriping with deblurring is superior to that handling each task individually. Some of the techniques proposed in the literature jointly considered deblurring and unmixing \cite{zhao2013deblurring, henrot2014does}. 
Henrot \emph{et al}. \cite{henrot2014does} derived a joint blurring observation and mixing model and showed how blur affects endmember identifiability within the geometrical unmixing framework.
Zhang \emph{et al}. \cite{zhang2012bayesian} proposed a Bayesian-based restoration approach using the fusion of hyperspectral and multispectral images to take into account the joint statistics of the datsaets. The estimation problem first divided into a \textcolor{black}{deblurring} and a denoising problem and they were iteratively solved using expectation minimization. 

\subsubsection{Blind Deblurring}
Depending on the prior knowledge of the PSF, the HSIs deblurring methods can be further divided into blind and nonblind HSI deblurring methods. Most of the existing HSI deblurring methods mainly focus on nonblind deblurring, i.e., the PSF is known while in real cases the blur kernels are  not known in advance. The blind HSI deblurring is much more difficult since both the HSI and blur kernel are unknown and should be estimated. Shen \emph{et al}. \cite{shen2012blind} proposed a blind image restoration method for the deblurring of remote sensing images, in which the Huber-Markov prior is employed to regularize both the clean image and blur kernel. 
Berisha \emph{et al}. \cite{berisha2015deblurring} considered the atmospheric turbulence deblurring of multiple PSF cases using a preconditioned alternating direction method of multipliers.

\subsubsection{Real-Time Application}
Due to the large size of HSIs, the running time of deblurring algorithms is another important issue. There are many works aiming at reducing running time for real-time applications. Li \emph{et al}. \cite{li2013image} proposed to perform deconvolution filter computations in the same support as the PSF, so that large matrix manipulations are avoided without memory limitations. Song \emph{et al}. \cite{song2019online} derived a sliding-block zero-attracting least mean square algorithm allowing the fast slice-by-slice online HSIs deconvolution. Avagian \emph{et al}. \cite{avagian2019fpga} implemented the Lucy-Richardson HSI deconvolution algorithm accelerated by FPGA. Song \emph{et al}. \cite{song2016regularization} improved the practicality of HSI deblurring methods by automatically estimating the regularization via the  minimum distance criterion (MDC) and maximum curvature criterion (MCC).

\begin{figure*}[htbp]
\begin{center}
    \includegraphics[width=0.99\textwidth]{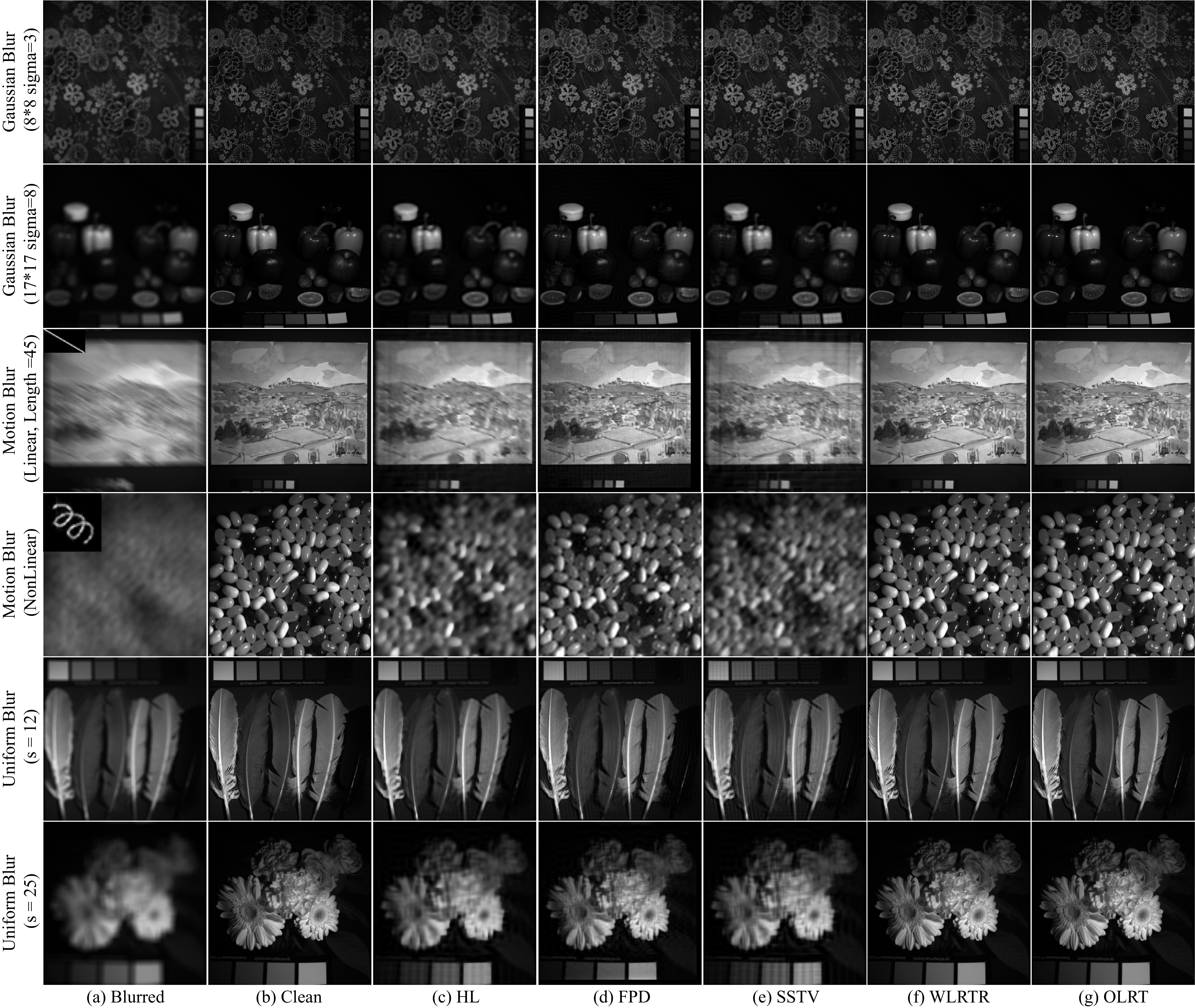}
 \end{center}
   \caption{Simulated deblurring results on the CAVE dataset. From the top to the down rows, we show the different blur cases and blur levels. From the left to the right columns, we show the deblurring results of the comparison methods. }
\label{Deblurring results}
\end{figure*}

\subsection{Results and Comparisons}
For HSIs deblurring, the competing methods include single-image-based deblurring method hyper-Laplacian (HL) \cite{krishnan2009fast}, gradient-based HSIs deblurring methods fast positive deconvolution (FPD) \cite{henrot2012fast}, deconvolution spectral-spatial TV  (SSTV) \cite{fang2017hyperspectral}, low-rank-based HSI deblurring methods weighted low-rank tensor recovery (WLRTR) \cite{chang2020weighted}, and optimal low-rank tensor method (OLRT) \cite{chang2020hyperspectral}. 
Note that most of the HSI deblurring methods are non-blind, i.e., the blur kernels are known in advance. 
\textcolor{black}{We apply the 3D convolution between the clean image and blur kernel to obtain the blurred HSIs. The 3D convolution is performed in the frequency domain, while the image is transformed by the MATLAB function ``fftn'' and the blur kernel is transformed by the MATLAB function ``psf2oft''. To generate different blur kernels, we employ the Matlab function ``fspecial'' to simulate the Gaussian blur, motion blur, and uniform blur with different blur degrees.} 

\subsubsection{Quantitative Evaluation} In Table \ref{CAVE_Deblurring Quantitative}, the deblurring results of the competing methods on different blur kernels and levels are shown. We choose three typical blurs: Gaussian blur, motion blur, and uniform blur. Each blur contains two different blur levels. 
The outcomes of the experiments can be \textcolor{black}{summarized} as follows: First, the low-rank tensor-based deblurring methods consistently obtain better results compared with the gradient sparsity-based methods. That is to say, the tensor-based methods could well preserve the 3D structure of the HSIs. Second, compared with the Gaussian or uniform blur, the motion blur is much easily restored. This is reasonable since the motion blur degrades the high-frequency image content only in a single direction while the Gaussian or uniform blur degrades high-frequency information in all directions. Third, the multi-frame-based methods usually perform better than that of the single image-based method HL \cite{krishnan2009fast}. That is to say, the utilization of the additional spectral information would be beneficial for HSI deblurring. 

\textcolor{black}{In Table \ref{Running Time Deblurring}, we report the running times of different deblurring methods. We perform the experiments on MATLAB 2017a, with an Intel i7 CPU at 3.6 GHz, an NVIDIA GEFORCE RTX 1080Ti GPU, and 32-GB memory. The HL is significantly faster than the other methods. The WLRTR and OLRT are computationally expensive due to the nonlocal cubic searching and higher-order SVD operation.} 

\subsubsection{Qualitative Evaluation} Fig. \ref{Deblurring results} shows the visual deblurring results of the competing methods. For each blur kernel case and blur level, we choose one typical result as representative. From Fig. \ref{Deblurring results}(a), it can be seen that the visual appearances of different blurs are obviously different. The single image-based HL can partially improve the visual effects and restore the sharp edge when the blur level is moderate. However, when the blur level is high, the single-image-based HL fails to obtain satisfactory results. As for the gradient-based HSI deblurring methods, FPD and SSTV restore better the high-frequency component. However, there are obvious ringing artifacts remaining in the results. The low-rank tensor-based deblurring methods have achieved impressive restoration results, in which both the sharp edge and texture are well recovered. Overall, the tensor-based methods seem to be most suitable for HSIs deblurring. 


\begin{table}[t]
\centering
\textcolor{black}{\caption{Processing time (in seconds) of the HSIs deblurring methods.}
\renewcommand{\arraystretch}{1.1}
\label{Running Time Deblurring}
\begin{tabular}{c|c|c|c|c|c}
\hline
      & HL & FPD & SSTV & WLRTR & OLRT \\ \hline
Times (Seconds) & 7  & 207 & 157  & 1421  & 2057 \\ \hline
\end{tabular}}
\end{table}

\subsubsection{\textcolor{black}{Discussion}} 
\textcolor{black}{The results of the low-rank tensor recovery methods are consistently better than those of other methods. This phenomenon has also been widely observed in HSIs denoising and destriping, which indicates two important aspects for general HSIs restoration. On the one hand, the tensor-based methods can well preserve the 3D structure of the HSIs. On the other hand, spectral correlations and spatial non-local self-similarities are the key ingredients to boost the final performance. However, the low-rank tensor methods usually require a very long processing time, as shown in Table \ref{Running Time Deblurring}. Moreover, it is worth noting that only few deep learning-based methods have been proposed yet for HSIs deblurring. We expect that CNN-based deblurring methods could further advance this field.}





\section{Summary, Conclusion, and Future Challenges}


Common methodologies have fueled the development of restoration methods in different fields of remote sensing. Variational methods, in particular, have offered for a long time a generic way to approach denoising, inpainting, and deblurring problems. While the data-fidelity term needs to be adapted to the specificities of each sensor/modality, this review shows that similar regularization terms have successfully been used in SAR imaging and HSI. TV and its extensions to multi-channel images have been widely employed for its edge-preserving property. Wavelet transform and sparsity constraints are two versatile concepts well-suited to remote sensing images, as they led to numerous developments these last twenty years. Patch-based processing has renewed the interest for improved filtering approaches and led to data-driven sparse coding techniques that learn a dictionary of patterns encountered in remote sensing images at the scale of small patches (typically about $8\times 8$ pixels). Since the advent of deep learning techniques, there is no doubt that these models offer unprecedented versatility. They push several steps further the natural evolution of image restoration techniques toward the learning of rich models directly from the data. Despite their very different physical principles, SAR and hyperspectral share several common properties. We believe that it is very enriching to analyze the challenges, successes, and trends faced when adapting deep learning models to remote sensing image restoration.

Like the field of bio-medical imaging, remote sensing faces a huge imbalance with, on the one hand, a gigantic amount of images available and, on the other hand, limited access to ground truth data. This challenge calls for several answers: (i) efforts by the scientific community to put forth high-quality databases to help advance research on deep learning techniques adapted to the restoration of remote sensing images; (ii) strategies to inexpensively fine-tune a network on a new sensor; (iii) the development of self-supervised techniques, i.e., techniques that rely solely on noisy data to learn a model of remote sensing scenes and of the degradation process. As one of the ongoing activities, the IEEE image analysis and data fusion technical committee (IADF) together with its partners have been organizing several activities (such as the data fusion contest of 2020 \cite{9295450} and 2021) to promote the use of noisy low-quality training data to produce high-quality classification and change maps.

Beyond their spatial dimensions, remote sensing images have additional dimensions: a spectral dimension in HSI, a polarimetric and/or interferometric dimension in SAR, and possibly a temporal dimension when time-series are considered. Deep neural networks have the ability to learn patterns from the data. However, when the number of dimensions of the images increases, the amount of training data required to cover all possible patterns that occur in multivariate remote sensing images explodes. Special attention must, thus, be paid to the network architecture in order to limit the degrees of freedom while keeping sufficient flexibility to capture complex patterns.

There are several research directions that we believe will play a significant role in the near future. The first direction is the exploitation of the huge image archives available for several satellites. Using deep neural networks, especially if trained without supervision, very rich models could be learned from all these images. The training strategies used to date only consider patches extracted from a collection of a few tens of images. Processing images at a global scale, covering several years/decades, represents a major leap. Given the computational cost of such training efforts, we hope that the trained networks obtained could be shared to the scientific community.
The capability of the models learned to generalize to other sensors is crucial. Transfer learning, domain adaptation techniques, or other strategies may be used in order to maximize the reusability of models trained on a given sensor.
Imaging satellites give access to the time-series of a scene. This temporal dimension offers powerful ways to restore the images by mitigating independent noise fluctuations. However, this requires separating fluctuations due to noise and fluctuations due to actual changes in the scene. Multi-temporal restoration of remote sensing images is an important research direction that calls for adequate network architectures.
Another direction is the cooperation between different sources of data: sensors with different spatial resolution, spectral coverage or radar band, incidence angle, or modality (e.g., SAR+HSI).
The interaction between deep neural networks and models of the physics of imaging (prior knowledge on the instrument, physical constraints that some parameters must fulfill) is yet another important research axis. 
Combining the expressivity of deep neural networks and domain-specific knowledge might be the key to the successful application of deep learning to cases with limited training data and to reinforce the confidence in the network outputs. Beyond the plausibility of the images produced by deep neural networks, it is also essential for the subsequent scientific exploitation of restoration results to characterize the reliability of the estimations. An analysis of the uncertainties of the results of deep neural networks is therefore a key question that may find answers in the Bayesian models that have been widely used in the field for several decades. 
Finally, self-supervision appears to be an extremely powerful training strategy for remote sensing, while seeming to still be at its infancy: some rather crude approaches have been applied, such as the deep image prior (which is computationally costly and requires an early stopping mechanism to prevent noise propagation), cycle-consistent adversarial strategies (which learn plausible image styles), or adding more noise to an already corrupted image (which obviously only works in sufficiently high signal-to-noise ratio regimes). Models based on pixel masking or on architectures with a receptive field that excludes a central area (named blind-spot) are promising. They can be trained either with a Bayesian framework that includes a statistical model of the noise, or in a noise-agnostic fashion. We expect more development of these approaches to emerge in the coming years.

In contrast to restoration approaches based on the inversion of a pre-determined model of the data, deep learning is flexible enough to adapt to pre-processed data, for example, ground range detected SAR images. This extends the range of application of restoration algorithms from low-level data, for which instrumental effects can be modeled, to high-level data. Thus, it bridges the gap between signal processing practices (getting back to the raw data to apply instrumental and statistical models) and end-users practices (use heavily pre-processed data, e.g., calibrated and ortho-rectified images). We believe that this is a major change that will benefit all thematic applications of remote sensing.



\section{Acknowledgements}
The work of Behnood Rasti was funded by the Alexander-von-Humboldt-Stiftung/foundation.


\ifCLASSOPTIONcaptionsoff
  \newpage
\fi




%

%
\bibliographystyle{IEEEbib}
\bibliography{references}

%








\end{document}